\documentclass[a4paper,11pt]{article}
\usepackage{jheppub}

\usepackage[T1]{fontenc}
\usepackage{bm}
\usepackage{amsmath}
\usepackage{amsfonts}
\usepackage{amssymb}
\usepackage{mathtools}
\usepackage{mathrsfs}

\renewcommand{\Re}{\mathop{\mathrm{Re}}}

\title{Two-loop kite master integral for a correlator of two composite vertices}

\author[a]{S.~V.~Mikhailov}
\author[a,b]{and N.~Volchanskiy}

\affiliation[a]{Bogoliubov Laboratory of Theoretical Physics, JINR,\\
					Joliot-Curie 6, 141980 Dubna, Russia}
\emailAdd{mikhs@theor.jinr.ru}
\affiliation[b]{Research Institute of Physics, Southern Federal University,\\
                Prospekt Stachki 194, 344090 Rostov-na-Donu, Russia}
\emailAdd{nikolay.volchanskiy@gmail.com}

\abstract{%
We consider the most general two-loop massless correlator $I(n_1,n_2,n_3,n_4,n_5;\allowbreak x,y;D)$ of two composite vertices with the Bjorken fractions $x$ and $y$ for arbitrary indices $\{n_i\}$ and space-time dimension $D$; this correlator is represented by a ``kite'' diagram. The correlator $I(\{n_i\};x,y;D)$ is the generating function for any scalar Feynman integrals related to this kind of diagrams. We calculate $I(\{n_i\};x,y;D)$ and its Mellin moments in a direct way by evaluating hypergeometric integrals in the $\alpha$ representation. The result for $I(\{n_i\};x,y;D)$ is given in terms of a double hypergeometric series---the Kamp\'{e} de F\'{e}rriet function. In some particular but still quite general cases it reduces to a sum of generalized hypergeometric functions $_3F_2$. The Mellin moments can be expressed through generalized Lauricella functions, which reduce to the Kamp\'{e} de F\'{e}rriet functions in several physically interesting situations. A number of Feynman integrals involved and relations for them are obtained.
}

\begin{document}

\maketitle

%================================================================
%================================================================
%================================================================

\section{\label{intro}Introduction}

The correlators of composite vertices appear naturally as the result of ``factorization'' of
scales in hard processes, more precisely in the technical sense---due to contractions of the so called
 ``hard subgraphs'' of the corresponding diagrams. In particular,
 such a two-point correlator with one composite vertex appears at the contraction of V-V subgraphs of the
$\langle V(q_1)V(q_2)A(p) \rangle$ triangle diagram for the kinematics with hard momentum transfer $-q^2_1$, $-q_2^2$ $\gg$ $p^2=(q_1+q_2)^2$,
where $V=\bar{\psi}\gamma_\mu \psi$ and $A=\bar{\psi}\gamma_\rho \gamma_5 \psi$ are the vector and axial
fermion currents, respectively. These contractions of the triangle constitute
a theoretical basis of the factorization approach for the perturbative QCD calculations of the transition form factors
for the processes $\gamma^*(q_1) \gamma^*(q_2) \to \pi^0(p)$, where $\pi^0$ is a neutral pion and $\gamma^*$'s are virtual photons.

Here we consider the calculation of a more general object than the one just mentioned---the two-point massless  correlator $I$ of \textit{two composite} vertices, which is the normalized Fourier
transform of the correlator $\langle J_\mu(0) J_\nu(\text{z}) \rangle$ of two composite
fermion currents $J_\mu$ and $J_\nu$,
see \cite{Mikhailov:1988nz}.
\begin{figure}[htb]
\centering
\begin{gather*} %\label{eq:1.1}
	I(p;\{n_i\};x,y;D)
	= \begin{gathered}\includegraphics{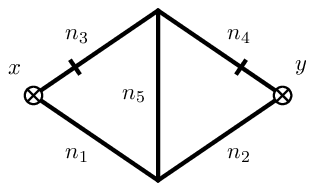}\end{gathered}
\end{gather*}
\caption{\label{fig:intro}Two-loop kite master integral for a correlator of two composite vertices $\otimes$ with fractions $x$ and $y$. The indices $n_i$, $i=1,\dots,5$ and the space-time dimension $D$ are arbitrary; $p$ is the external momentum. The slash on a line connected to a composite vertex with fraction $u=x,y$ designates a Dirac delta $\delta(u-\tilde{n}q)$ in the integrand, where $q$ is the momentum (outgoing w.r.t.\ the composite vertex) of a line, $\tilde{n}$ is a light-cone vector normalized so that $(\tilde{n}p) = 1$.}
\end{figure}
As it happens, a more general quantity can be evaluated technically easier than its superficially simpler counterpart with only one single composite vertex $J_\nu$.
Our goal is the calculation of the two-loop massless ``kite'' scalar diagram $I(p;\{n_i\};x,y;D)$ in figure~\ref{fig:intro},
taken at any values of indices $n_i$ of the lines,
$\{n_i\} = n_1,n_2,n_3,n_4,n_5$, and any space-time dimension $D$.
The ``kite'' diagram is one of the master integrals for the two-current correlator
at two and three loops, see eq.~(\ref{eq:2loopscint}).
The function $I$  is the generating function for any Feynman integral related to this kind of diagrams.
The integrals can be obtained by convolving the function $I$
with appropriate  weights $\varphi$ and $\phi$, $\varphi(x)\otimes I(p;\{n_i\};x,y;D)\otimes \phi(y)$,
where symbol $\otimes$ means integration over the longitudinal momentum fractions $x$ or $y$.
In the coordinate representation the weight function $\varphi$ (or $\phi$) becomes an operator $\varphi(\tilde{n}\cdot {\vec \partial}_\text{z}/(\tilde{n}\cdot p))$
that acts at the vertex point z ($\tilde{n}^2=0$).
To return to the correlator with one composite vertex, e.g.\ the one with the fraction $x$, we should integrate $I(p;\{n_i\};x,y;D)$ over the fraction $y$.
Besides, the original  two-argument correlator $I(\ldots;x,y;D)$ was used to analyse the properties of the conformal composite vertices under renormalization in \cite{Craigie:1983fb}.

The zeroth Mellin moments of $I(\ldots;x,y;D)$ in both $x$ and $y$,  $I(\ldots;\underline{0},\underline{0};D)$,\footnote{Throughout this paper, the Mellin transform of a function $f(x)$ is indicated by underscoring its argument, i.e. $f(\underline{a}) = \int_0^1 x^a f(x) \, \mathrm{d}x.$}
give the ordinary master integral of kite topology. It has been known for relatively long time that this integral can be evaluated in closed form involving hypergeometric functions. In ref.~\cite{Chetyrkin:1980pr}, Chetyrkin et al.\ showed for the first time that the integral $I(\ldots;\underline{0},\underline{0};D)$ can be expressed as a double hypergeometric series for $n_1 = n_2 = 1$. To come up with this result, they expanded the integrand (in coordinate space) over a basis of the Gegenbauer polynomials and successfully solved the remaining integrals. A special case $n_1 = n_2 = n_3 = n_4 = 1$ was considered by Kazakov \cite{Kazakov1985} and Broadhurst \cite{Broadhurst:1985vq} a few years later. They applied the methods of uniqueness and integration by parts (IBP) to derive functional equations for the integral under consideration and then solved them in terms of the function $_3F_2(-1)$\footnote{We use sometimes the designation $_pF_q(x)$ with omitted parameters implying that the values of the parameters are irrelevant in the particular context or left generic.}. Another result in terms of $_3F_2(1)$ was obtained by Kotikov who refined the technique of ref.~\cite{Chetyrkin:1980pr} and found a transformation from the double series to a single one even in a more general case \cite{1996PhLB..375..240K}. Recently, the equivalence of the representations through $_3F_2(\pm 1)$ was proved in ref.~\cite{Kotikov:2016rgs}. In refs.~\cite{Barfoot:1987kg} and \cite{Broadhurst:1996ur}, several different representations in terms of $_4F_3(1)$ and $_3F_2(1)$ were obtained by cleverly using the symmetries of the integral \cite{1985TMP....62..232G,Broadhurst:1986bx,Barfoot:1987kg}. For more details and complete lists of references we suggest reading reviews \cite{2012IJMPA..2730018G,2018arXiv180505109K} and the book by Grozin \cite{Grozin:2007}. In this work, we rely on a different approach to derive the general results---a direct calculation of hypergeometric integrals occurring in the $\alpha$ representation as suggested in \cite{Mikhailov:1988nz} for the special case $D=4$ (see especially the preprint \cite{Mikhailov:1988nz-JINRrep} for technical details).

The most general case of $I(\ldots;\underline{0},\underline{0};D)$ with arbitrary indices was studied in \cite{Bierenbaum2003}. There, the corresponding Feynman integral was calculated in the Mellin--Barnes representation and expressed in terms of double hypergeometric series (for earlier less general results obtained in the same fashion for $D=4$ see \cite{belokurov1984calculating6082, Usyukina1989}). Acting essentially in the same way, we obtain the twofold Mellin moment $I(\ldots;\underline{a},\underline{N};D)$ with $N$ and $a$ being a natural number and a real one, respectively. The result is a sum of the Kamp\'{e} de F\'{e}rriet (KdF) functions. Even more general case of the twofold Mellin moment $I(\ldots;\underline{a},\underline{b};D)$ with an arbitrary real $a$ and $b$ is worked out in terms of the generalized Lauricella functions of three variables (more precisely, the series $F^{(3)}(z_1,z_2,z_3)$ introduced by Srivastava \cite{srivastava1967generalized}; see also section~1.5 in \cite{srivastava1985multiple}).

Evaluating master integrals in hypergeometric series is of value for at least two reasons. First of all, it opens up one of the avenues to expose analytic properties of the integrals using the machinery that has been developed in the theory of special functions. Secondly, in many physically relevant cases we can almost immediately expand these multivariate functions in dimensional regularization parameter. To this end, one can employ the method developed by Moch, Uwer, and Weinzierl some time ago \cite{2002JMP....43.3363M} and suitable for expansions of KdF, Lauricella, and generalized hypergeometric functions in the vicinity of an even number of space-time dimensions (\texttt{C++} and \texttt{FORM} implementations of the method are described in \cite{2002CoPhC.145..357W, Moch:2005uc}). For univariate functions $_pF_q$ there are a variety of algorithms and general theorems about the Laurent expansion of these functions near even and odd $D$, e.g.\ \cite{2006JHEP...04..056K, 2007JHEP...11..009K, 2007JHEP...02..040K, Huber:2005yg, 2008CoPhC.178..755H} (see also references therein).

The rest of the paper is organized as follows. In section \ref{sec:one-loop}, we introduce our notation and illustrate it on the simplest one-loop example of the two-current correlator. In section~\ref{subsec:method}, we discuss in detail our method of evaluating the integral $I(\ldots;x,y;D)$ that results in a representation for $I$ in terms of the KdF functions of two variables in section~\ref{subsec:KdF}. Section~\ref{subsec:3F2} is devoted to some important cases of reduction of these KdF functions to hypergeometric series $_3F_2$. In section~\ref{sec:MMG-gen}, the calculation of various moments of $I$ is discussed. We consider the  case of the indices  $\{1,1,1,1, n_5=n\}$, which is  important for QCD applications, in section~\ref{sec:special}. There we also derive Mellin moments of $I(p;1,1,1,1,n; x,y;D)$ and compare them with the results obtained in the literature previously. Our conclusions are given in section~\ref{sec:concl}, while the definitions of the hypergeometric functions used in this paper and some technical issues are treated in three appendices.

%===================================================
%===================================================
%===================================================

\section{\label{sec:one-loop}Simple example: one-loop integral}

First, we introduce a generalization of the $G$ function for the one-loop integral with composite vertices $\otimes$:
\begin{align} \label{eq:1lscint}
	I(p;n_1,n_2;x,y;D)
	&= \begin{gathered}\includegraphics{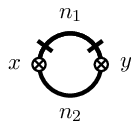}\end{gathered}
=
	\delta\left(x-y\right) \int \frac{\mathrm{d}^D k \,\delta\left(x-\tilde n k \right)}{[k^2]^{n_1} [(k-p)^2]^{n_2}}
\notag\\
&=
	(-)^{n_1+n_2} i \pi^{D/2} \left(-p^2\right)^{D/2 - n_1-n_2} \delta\left(x-y\right)  G(n_1,n_2;x;D),
\end{align}
where a slash (beside the composite vertex) on the line with momentum $k$ means factor $\delta(x-\tilde{n}k)$; $\tilde n_\mu$ is a light-cone vector, $\tilde n^2 = 0$, normalized so that $(\tilde np) = 1$. The function $G(n_1,n_2;x;D)$ is dimensionless and reduces to the usual one-loop $G$ function (see appendix B of ref.~\cite{Chetyrkin:1980pr}, section~1.5 in \cite{Grozin:2007} or section~3.1 in \cite{Smirnov:2006}) if it is integrated over $x$:
\begin{gather}
	\int_0^1 \mathrm d x \, G(n_1,n_2;x;D) = G(n_1,n_2)
	=
	\frac{(-)^{-n_1-n_2}}{i \pi^{D/2}} \left(-p^2\right)^{n_1+n_2-D/2} \int \frac{\mathrm{d}^D k}{[k^2]^{n_1} [(k-p)^2]^{n_2}}.
\end{gather}
The function $G(n_1,n_2;x;D)$ has obvious symmetry:
\begin{gather}
	G(n_1,n_2;x;D) = G(n_2,n_1;\bar x;D), \qquad \bar x = 1-x.
\end{gather}

The integral \eqref{eq:1lscint} is easily calculated when the propagators $1/[k^2]^{n_1}$ and  $1/[(k-p)^2]^{n_2}$ are cast into the $\alpha$ representation and the Dirac delta function is substituted by its Fourier integral.
The result for the $G$ function reads
\begin{gather}\label{eq:Gx1loop}
	G(n_1,n_2;x;D)
	=
	\frac{\Gamma(n_1+n_2-D/2)}{\Gamma(n_1)\Gamma(n_2)} 	x^{D/2-n_1-1} \bar x^{D/2-n_2-1}.
\end{gather}
The Mellin transform of the function $G$ is
\begin{align}
	G(n_1,n_2;\underline{a};D) &{}= \int_0^1 \mathrm d x \, x^a G(n_1,n_2;x;D)
	= \Gamma\begin{bmatrix}n_{1,\tilde 2},\; a-n_{\tilde 1},\; -n_{\tilde 2} \\ n_1,\; n_2,\; a-n_{\tilde 1, \tilde 2}\end{bmatrix}
\notag\\ \label{eq:1loopG}
&{}=
\frac{\Gamma(n_1+n_2-D/2)\Gamma(D/2-n_1+a)\Gamma(D/2-n_2)}{\Gamma(n_1)\Gamma(n_2)\Gamma(D-n_1-n_2+a)}.
\end{align}
Here and in what follows, a $\nu$th Mellin moment of a function $f(x)$ is denoted by the underlined argument of the function $f(\underline\nu)$; the $n$'s with multi-indices are defined as
\begin{gather}
	n_{i_1,\dots i_K, \tilde j_1, \dots \tilde j_L} = \sum^K_{k=1} n_{i_k} + \sum^L_{k=1} (n_{j_k}-D/2).
\end{gather}
The definition of the two-row $\Gamma$ function is clear from eq.~\eqref{eq:1loopG}:
\begin{gather}
	\Gamma\begin{bmatrix}a_1,\dots,\; a_p \\ b_1,\dots,\; b_q\end{bmatrix}
	= \frac{\prod_{i=1}^{p} \Gamma(a_i)}{\prod_{i=1}^{q} \Gamma(b_i)}.
\end{gather}

A slightly more general one-loop integral with $(\tilde n q) \neq 1$ also comes in handy:
\begin{align}
	&I(p,q;n_1,n_2;x;D)
	=
	\int \frac{\mathrm{d}^{D} k \,\delta\left(x-\tilde n k\right)}{[k^2]^{n_1} [(k-q)^2]^{n_2}}
\notag\\
	&{}=
	(-)^{n_{1,2}} i \pi^{D/2} \left(-q^2\right)^{D/2 - n_{1,2}}
	\int_0^1 \mathrm{d} y \, G\left(n_1,n_2;y;D\right) \delta \left[ x-y (\tilde n q) \right]
\notag\\
	&{}=
	(-)^{n_{1,2}} i \pi^{D/2} \left(-q^2\right)^{D/2 - n_{1,2}} G\left(n_1,n_2;\frac{x}{(\tilde n q)};D\right) \frac1{(\tilde n q)} \Theta\left( \frac{x}{(\tilde n q)} \right) \Theta\left( 1- \frac{x}{(\tilde n q)} \right),
\label{eq:1lscint2}\end{align}
where $\Theta(x)$ is the Heaviside step function.

%=============================================================
%=============================================================
%=============================================================

\section{\label{sec:general}Kite-type correlator as a generalized hypergeometric series}

%=============================================================
%=============================================================
%=============================================================

\subsection{\label{subsec:method}The correlator in the \boldmath$\alpha$ representation. Reduction to a double integral}

Let us now consider a general two-loop master integral with composite external vertices:
\begin{align}
	I(p;\{n_i\};x,y;D)
		&{}= \begin{gathered}\includegraphics{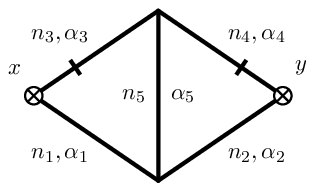}\end{gathered}
\notag\\
&{}=	\int \frac{\mathrm{d}^D k_1 \, \mathrm{d}^D k_2}{\prod_{r=1}^5 \mathscr{D}_r^{n_r}} \,\delta\left(x-\tilde nk_1\right)  \delta\left(y-\tilde nk_2\right)
\notag\\
\label{eq:2loopscint}
	&{}=
	(-)^{ n_{1,2,3,4,5}+1} \pi^{D} \left(-p^2\right)^{\omega/2} G(\{n_i\};x,y;D),
\end{align}
where $\omega = -2 n_{1,2,3,\tilde 4,\tilde 5} = 2D - 2 \sum n$ is the degree of divergence of the integral and the propagator factors $\mathscr{D}_k$ in the denominator are
\begin{align}\label{eq:denoms}
	\mathscr{D}_1 = (k_1-p)^2,
	\qquad
	\mathscr{D}_2 = (k_2-p)^2,
	\qquad
	\mathscr{D}_3 = k_1^2,
	\qquad
	\mathscr{D}_4 = k_2^2,
	\qquad
	\mathscr{D}_5 = (k_1-k_2)^2.
\end{align}

The integral \eqref{eq:2loopscint} is symmetric under the following interchanges of its parameters:
\begin{gather}\label{eq:2loopsym1}
	G(n_1,n_2,n_3,n_4,n_5;x,y;D)
	=
	G(n_2,n_1,n_4,n_3,n_5;y,x;D),
	\\\label{eq:2loopsym2}
	G(n_1,n_2,n_3,n_4,n_5;x,y;D)
	=
	G(n_3,n_4,n_1,n_2,n_5;\bar x, \bar y;D).
\end{gather}

If one of the indices $n_r$ vanishes, the integral \eqref{eq:2loopscint} is a convolution of the one-loop integrals \eqref{eq:1lscint}. Evaluating \eqref{eq:2loopscint} with the help of the one-loop integrals \eqref{eq:Gx1loop} and \eqref{eq:1lscint2}, we indeed obtain
\begin{align}\label{eq:G2n_5=0}
	G(n_1,n_2,n_3,n_4,0;x,y;D)
	&{}=
	G(n_3,n_1;x;D) G(n_4,n_2;y;D),
\\
	G(0,n_2,n_3,n_4,n_5;x,y;D)
	&{}=
	  \frac{\Theta(y-x)}{y} G\left(n_3,n_5;\frac{x}{y};D\right) G(n_{3,4,\tilde 5},n_2;y;D)
	\notag\\
	&\mspace{-80mu}{}= \Theta(y-x) \, x^{-n_{\tilde 3}-1} (y-x)^{-n_{\tilde 5}-1}
	y^{-n_4} \bar y^{-n_{\tilde 2}-1}
	\Gamma\begin{bmatrix}n_{3, \tilde 5},\; n_{2,3,\tilde 4, \tilde 5} \\ n_2, \; n_3,\; n_5,\; n_{3,4,\tilde 5} \end{bmatrix},
\label{eq:0nnnn}
\\
	G(n_1,0,n_3,n_4,n_5;x,y;D)
	&{}=
	G(0,n_1,n_4,n_3,n_5;y,x;D),
\\
	G(n_1,n_2,0,n_4,n_5;x,y;D)
	&{}=
	G(0,n_4,n_1,n_2,n_5;\bar x, \bar y;D),
\\\label{eq:G2n_4=0}
	G(n_1,n_2,n_3,0,n_5;x,y;D)
	&{}=
	G(0,n_3,n_2,n_1,n_5;\bar y,\bar x;D).
\end{align}
Note that the $(\underline 0, \underline 0)$th Mellin moment of eq.~\eqref{eq:0nnnn} is 0th moment of the Mellin convolution of two one-loop functions that is a product of 0th moments of these functions, which is a well known property of the two-loop master integral $G(\{n_i\};\underline 0,\underline 0;D)$ (section~4.1 in \cite{Grozin:2007}).

In the cases of $n_1=n_3=0$ and $n_2=n_4=0$, the integral \eqref{eq:2loopscint} vanishes---it reduces to a product of one-loop integral and vacuum loop that is zero in dimensional regularization (at least for $D \neq 2 n_5$):
\begin{gather}
	I(p;0,n_2,0,n_4,n_5;x,y;D) = \delta(x-y) I(p;n_4,n_2;y;D) I(0;n_5,0;\underline 0 ; D),
	\\
	I(p;n_1,0,n_3,0,n_5;x,y;D) = \delta(x-y) I(p;n_3,n_1;x;D) I(0;n_5,0;\underline 0 ; D),
\end{gather}
where $I(0;n_5,0;\underline 0 ; D)$ is the one-loop vacuum integral, which was considered in ref.~\cite{1985TMP....62..232G}:
\begin{gather}
	I(0;n,0;\underline 0 ; D) = \int \frac{\mathrm d^D\, k}{(k^2)^n} \sim \delta(D-2n).
\end{gather}

To evaluate the integral \eqref{eq:2loopscint} in the general case of all $n_r$ being non-zero, we borrow a trick invented in ref.~\cite{Mikhailov:1988nz-JINRrep} (see appendix A therein). Firstly, we substitute the denominators $\mathscr D_r^{n_r}$ by their $\alpha$ representation and the Dirac delta functions by their Fourier integrals. As usual, this allows us to evaluate momentum integral as the Gaussian one, but also leaves a trace of two Dirac delta functions. We have
\begin{multline}\label{eq:I1}
	I(p;\{n_i \};x,y;D)
=
	(-)^{n_{1,2,3,4,5} +1} \pi^{D} \int_{0}^{\infty} \prod_{r=1}^5 \left[ \frac{\mathrm{d} \alpha_r \, \alpha_r^{n_r-1}}{\Gamma(n_r)} \right] e^{p^2 A_0/\mathscr D}
\\
	{}\times
 \delta\left( x-\frac{A_1}{\mathscr D} \right) \delta\left( y-\frac{A_2}{\mathscr D} \right) \frac{1}{\mathscr D^{D/2}}.
\end{multline}
Here, $\mathscr D$, $\lambda_s$, $s=1$, 2, and $A_s$, $s=0$, 1, 2, are the Symanzik polynomials of the parameters $\alpha$,
see figure in eq.~(\ref{eq:2loopscint}), and the fractions $x$ and $y$:
\begin{align}
\begin{alignedat}{2}
	\mathscr{D} &{}= \alpha_{1,3,5} \alpha_{2,4,5} - \alpha_5^2,
& \qquad
	A_0&{} = \alpha_3 A_1 + \alpha_4 A_2,\;\;
\\
	A_1 &{}= \alpha_1 \alpha_{2,4,5} + \alpha_2 \alpha_5,
&
	A_2 &{}= \alpha_2 \alpha_{1,3,5} + \alpha_1 \alpha_5.
\end{alignedat}
\end{align}

Secondly, we apply the Borel transform to both sides of eq.~\eqref{eq:I1}. The definition of the Borel transform and the necessary images are as follows:
\begin{gather}
	B \left[ f(t) \right](\mu) = \lim_{\begin{subarray}{c} t=n\mu \\ n\to \infty\end{subarray}} \frac{(-t)^n}{\Gamma(n)} \frac{\mathrm d^n }{\mathrm d t^n}f(t),
\\
	B \left[ e^{-at} \right](\mu) = \delta(1-\mu a), \quad a >0,
\qquad
	B \left[ t^{-a} \right](\mu) = \frac{\mu^{-a}}{\Gamma(a)}, \quad a >0.
\end{gather}
Upon substituting the left-hand side of \eqref{eq:I1} by the last line in eq.~\eqref{eq:2loopscint}, we can apply the Borel transform ($-p^2 \to M^2$) to both sides of the resulting equality and rescale the parameters $\alpha$ by the Borel parameter $M^2$, $\alpha_r \to \alpha_r/M^2$. Eventually, we arrive at the following expression:
\begin{multline}\label{eq:I3}
	G(\{n_i \};x,y;D)
=
	\Gamma\left(-\frac{\omega}{2}\right) \int_{0}^{\infty} \prod_{r=1}^5 \left[ \frac{\mathrm d \alpha_r \, \alpha_r^{n_r-1}}{\Gamma(n_r)} \right]
\\
	{}\times \delta\left( x-\frac{A_1}{\mathscr D} \right) \delta\left( y-\frac{A_2}{\mathscr D} \right) \delta\left( 1-\frac{A_0}{\mathscr D} \right) \frac{1}{\mathscr D^{D/2}} .
\end{multline}
Finally, enjoying the plentiful amount of the Dirac deltas, we can immediately eliminate three integrals. To this end, it is convenient to make substitutions
\begin{gather}
	\alpha_1 = \frac{a}{\bar x}, \qquad \alpha_2 = \frac{e}{\bar x}, \qquad \alpha_3 = \frac{b}{\bar x}, \qquad \alpha_4 = \frac{c}{\bar x}.
\end{gather}
Then, integrating over $\alpha_5$, $c$, and $e$, we obtain a factorized expression
\begin{multline}\label{eq:I4}
	G(\{n_i \};x,y;D)
=
	\frac{\Gamma\left(-\omega/2\right)}{\prod_{r=1}^5 \Gamma(n_r)}
	 \bar x^{-1-n_{\tilde 1}} \bar y^{-1-n_{\tilde 2}}  x^{-1-n_{\tilde 3}} y^{-1-n_{\tilde 4}} \lvert x-y \rvert^{-n_5}
 \\
	{}\times \left[ \Theta(\bar z) F(n_1,n_2,n_3,n_4,n_5;z;D) + \Theta(-\bar z) F(n_3,n_4,n_1,n_2,n_5;1/z;D)\right] .
\end{multline}
Here, $F(n_1,n_2,n_3,n_4,n_5;z;D)$ is a function of $z=(y \bar x) /(x \bar y)$ and is defined as a leftover double integral:
\begin{gather} \label{eq:Fz}
	F(\{n_i \};z;D) = \bar z^\lambda \int_0^1 \mathrm d a \int_0^a \mathrm d b\, \frac{a^{n_1-1} \bar a^{n_2-1} b^{n_3-1} \bar b^{n_4-1} (a-b)^{n_5-1}}{ \left(a \bar b -\bar a b z\right)^\lambda},
\\ \label{eq:Fsym}
	F(n_1,n_2,n_3,n_4,n_5;z;D) = F(n_4,n_3,n_2,n_1,n_5;z;D),
\end{gather}
where $\lambda=D/2 -1$. Note that the expression in the square brackets in eq.~\eqref{eq:I4} is a function of a single variable defined as a ratio of fractions $x$ and $y$,  conformal ratio $z=(y \bar x) /(x \bar y)$. We borrowed the name for $z$ from ref.~\cite{Braun2017}  (see also references therein), since the form of the ratio closely resembles that appearing in the evolution kernel for light-ray operator as a consequence of the conformal group.

%===================================================
%===================================================
%===================================================

\subsection{\label{subsec:KdF}The correlator as the Kamp\'{e} de F\'{e}riet function}

For arbitrary nonvanishing $n_r$ one of the integrations in eq.~\eqref{eq:Fz} can be easily performed in terms of the Appell function $F_1$. Indeed, making a substitution $b = a w$, we have as an integral over $w$ a classic Euler-type integral representation of the first Appell function \eqref{fwc:07.36.07.0001.01}:
\begin{align}
	F(\{n_i\};z;D)&{}=
	\Gamma\begin{bmatrix}n_3,\; n_5\\ n_{3,5}\end{bmatrix} \bar z^\lambda
	\int_0^1 \mathrm d a\, a^{n_{1,3,\tilde 5}-1} \bar a^{n_2-1}
	F_1 \left. \left( \begin{matrix}n_3;\,\lambda,\, 1-n_4\, \\ n_{3,5}\end{matrix} \right\rvert a +\bar a z, a  \right)
\notag\\ \label{eq:appellrep1}
	&{}=\Gamma\begin{bmatrix}n_3,\; n_5\\ n_{3,5}\end{bmatrix}
	\int_0^1 \mathrm d a\, a^{n_{1,3,\tilde 5}-1} \bar a^{n_{2,4,\tilde 5}-1}
	F_1 \left. \left( \begin{matrix}n_5;\,\lambda,\, n_{3,4,\tilde 5}\, \\ n_{3,5}\end{matrix} \right\rvert -\frac{z}{\bar z}, a \right).
\end{align}
The second equality in the equation above can be easily obtained by applying the autotransformation properties of the Appell function \eqref{fwc:07.36.17.0005.01}.

In the general case, the remaining integral over $a$ can be evaluated in terms of the Kamp\'{e} de F\'{e}riet (KdF) function $\mathsf{f}$
\begin{gather}\label{eq:FthKdF}
	F(\{n_i\};z;D)=
	\Gamma\begin{bmatrix}n_3,\; n_{2,4,\tilde 5} \\ \lambda,\, n_{3,4,\tilde 5} \end{bmatrix}
	\mathsf{f}^{1:1;2}_{1:0;1} \left( \left.\begin{matrix}n_5&:&\lambda&;&n_{3,4,\tilde 5},\,n_{1,3,\tilde 5}\,\\ n_{3,5}&:& \text{---} &;&n_{1,2,3,4,\tilde 5,\tilde 5}\,\end{matrix} \right\rvert -\frac{z}{\bar z}, 1 \right).
\end{gather}
The KdF function $\mathsf{f}$ and the Appell function $F_1$ are defined as hypergeometric series \eqref{eq:gen-KdF-def} and \eqref{eq:AppellF1}, respectively (see appendix~\ref{app:HGFs-defs}). With the help of eqs.~\eqref{eq:110f121-au} and \eqref{eq:110f121-au2}, we can also write the above KdF function in the form of series in a variable on the interval $[0,1]$ due to the step functions in \eqref{eq:I4}:
\begin{align}\label{eq:FthK-AC}
	F(\{n_i\};z;D)={}&
	\Gamma\begin{bmatrix}n_5,\, 1+n_2-n_3,\, n_3-n_2,\, n_{2,4,\tilde 5} \\ \lambda,\, n_{3,4,\tilde 5},\, n_{1,2,\tilde 5},\, n_{2,4,\tilde 5} \end{bmatrix} \bar{z}^{\lambda}
\notag \\
	&{} \times \Biggl\{ \mathsf{f}^{1:1;2}_{1:0;1} \left( \left.\begin{matrix}n_2&:&\lambda&;&n_{2,4,\tilde 5},\,n_{1,2,\tilde 5}\,\\ n_{2,5} &:& \text{---} &;&1+n_2-n_3\,\end{matrix} \right\rvert z, 1 \right)
\notag \\
	&\hphantom{\times \Biggl\{}{}-\mathsf{f}^{1:1;2}_{1:0;1} \left( \left.\begin{matrix}n_3&:&\lambda&;&n_{3,4,\tilde 5},\,n_{1,3,\tilde 5}\,\\ n_{3,5}&:& \text{---} &;&1+n_3-n_2\,\end{matrix} \right\rvert z, 1 \right)
	\Biggr\}
\notag \\
={}& \Gamma\begin{bmatrix}n_{2,4,\tilde 5} \\ \lambda,\, n_{3,4,\tilde 5} \end{bmatrix}
\bar{z}^\lambda \mathsf{f}^{0:2;3}_{1:0;1} \left( \left.\begin{matrix}\text{---}&:&n_{3},\,\lambda&;&n_5,\,n_{3,4,\tilde 5},\,n_{1,3,\tilde 5}\,\\ n_{3,5}&:& \text{---} &;&n_{1,2,3,4,\tilde 5,\tilde 5}\,\end{matrix} \right\rvert z, 1 \right).
\end{align}

%===================================================
%===================================================
%===================================================

\subsection{\label{subsec:3F2}Reduction to a univariate hypergeometric series}

We can isolate some cases when the complicated hypergeometric series \eqref{eq:FthKdF} reduces to a simpler one. The simplest instances of such reduction are for $n_5=0$ and $n_3=0$. In the former case the  KdF function \eqref{eq:FthKdF} is unity. In the latter one the  KdF function \eqref{eq:FthKdF} is a product of ${}_1F_0(\lambda,-z/\bar z) = \bar z^\lambda$ and $_2F_1(n_{3,4,\tilde 5},\,n_{1,3,\tilde 5},n_{1,2,3,4,\tilde 5,\tilde 5};1)$, which can be expressed through Euler gamma functions. It is easy to prove that we have the same results for the functions $G$ with one vanishing index as ones previously obtained in eqs.~\eqref{eq:G2n_5=0}--\eqref{eq:G2n_4=0}.

Another useful for calculations reduction is obvious from the Eulerian integral \eqref{eq:appellrep1}---the Appell double series is simplified to the hypergeometric function if we set $n_4=1$,
\begin{gather}
	F(n_1,n_2,n_3,1,n_5;z;D)
=
	\Gamma\begin{bmatrix}n_3,\; n_5\\ n_{3,5}\end{bmatrix} \bar z^\lambda
	\int_0^1 \mathrm d a\, a^{n_{1,3,\tilde 5}-1} \bar a^{n_2-1}
	{}_2F_1 \left. \left( \begin{matrix}n_3,\, \lambda \, \\ n_{3,5} \end{matrix} \right\rvert  a +\bar a z\right).
\end{gather}
Applying one of the Kummer transformations \eqref{fwc:07.23.17.0061.01} expressing $_2F_1(z)$ in terms of two $_2F_1(\bar z)$ and using the integral representation of $_3F_2$ \eqref{fwc:07.31.07.0001.01} result in
\begin{align}
	F(n_1,n_2,&n_3,1,n_5;z;D)
\notag\\
	={}&\Gamma\begin{bmatrix}n_{2,\tilde 5}+1,\, n_{1,3,\tilde 5},\, n_5,\, \lambda- n_{5}\\ n_{1,2,3,\tilde 5,\tilde 5}+1,\, \lambda \end{bmatrix} \bar z^{n_{ 5}}
	{}_3F_2 \left( \begin{matrix}n_5,\, n_{2,\tilde 5}+1,\, n_{3,\tilde 5}+1 \\ n_{\tilde 5}+2,\, n_{1,2,3,\tilde 5,\tilde 5}+1 \end{matrix} \right\rvert \left. \bar z \vphantom{\begin{matrix}n_3 \\ n_{3,5}\end{matrix}} \right)
\notag\\ \label{eq:KdFto3f2red1}
	&{}+ \Gamma\begin{bmatrix}n_3,\; n_{\tilde 5}+1,\; n_2,\; n_{1,3,\tilde 5}\\ n_{3,\tilde 5}+1,\; n_{1,2,3,\tilde 5} \end{bmatrix}
	{\bar{z}^\lambda} {}_3F_2 \left( \begin{matrix}\lambda, \; n_2,\; n_3 \\ -n_{\tilde 5},\; n_{1,2,3,\tilde 5} \end{matrix} \right\rvert \left. \bar z \vphantom{\begin{matrix}n_3 \\ n_{3,5}\end{matrix}} \right).
\end{align}

\begin{figure}[t]
\centering
\includegraphics{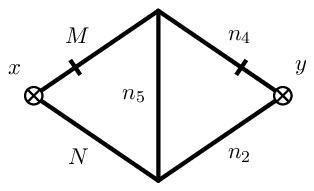}
\quad
\includegraphics{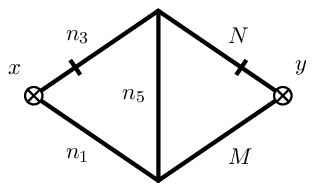}
\\
\includegraphics{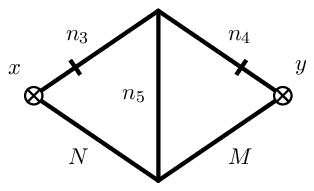}
\quad
\includegraphics{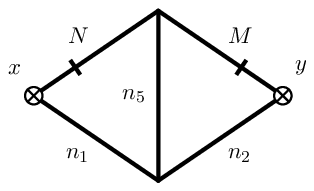}
\caption{\label{fig:condition}The cases that admit a reduction of the KdF function \eqref{eq:FthKdF} to a sum of generalized hypergeometric functions as given by eqs.~\eqref{eq:KdFto3f2red1} and \eqref{eq:KdFto3f2red2}. $N$ and $M$ are natural numbers.}
\end{figure}

For arbitrary $n_4$ we can derive a series representation for the function $F$ from eqs.~\eqref{eq:appellrep1} and \eqref{eq:AppellF1}:
\begin{gather} \label{eq:KdFto3f2red2}
	F(n_1,n_2,n_3,n_4,n_5;z;D)
=
	\sum_{r=0}^{\infty} \frac{\Gamma(1-n_4+r)}{r!\Gamma(1-n_4)} F(n_1,n_2,n_3+r,1,n_5;z;D).
\end{gather}
If $n_4$ is a natural number, $n_4=N$, the series truncates at $r=N-1$. Therefore, due to the symmetries \eqref{eq:2loopsym1}, \eqref{eq:2loopsym2}, and \eqref{eq:Fsym}, the function $G(n_1,n_2,n_3,n_4,n_5;x,y;D)$ as represented by eq.~\eqref{eq:I4} is a finite sum of hypergeometric functions $_3F_2$ if two adjacent external edges of the diagram have indices that are natural numbers (e.g., $n_1$, $n_2 \in \mathbb N$),
see figure~\ref{fig:condition}.

%==================================================================
%==================================================================
%==================================================================

\subsection{\label{sec:MMG-gen}Mellin moments of $G(\{n_i\};x,y;D)$}

One- and twofold Mellin moment of the correlator \eqref{eq:I4},
\begin{gather}\label{eq:MMG-gen-def}
	G(\{n_i\};x,\underline{b};D)
	=
	\int_0^1 \mathrm{d}y\, y^b G(\{n_i\};x,y;D),
\quad
	G(\{n_i\};\underline{a},\underline{b};D)
	=
	\int_0^1 \mathrm{d}x\, x^a G(\{n_i\};x,\underline{b};D),
\end{gather}
can be expressed in terms of the generalized Lauricella hypergeometric function. To prove this, we should simply make the following changes of integration variables: $y \to z=(y \bar x) /(x \bar y)$ in the above Mellin integrals \eqref{eq:MMG-gen-def} and $b \to c=(\bar a b)/(a \bar b)$ in \eqref{eq:Fz}. As a result, we obtain
\begin{align}
	G(\{n_i\};x,\underline{b};D)
	={}
	&x^{b-n_{3,\tilde 4,\tilde 5}-1} \bar x^{-n_{1,\tilde 2,\tilde 5}-1} \Gamma\left[P_1\right] \mathsf{f}^{0:3;3;2}_{2:1;0;0} \left( \left. Q_1 \right\rvert 1, 1, x \right){}+
\notag\\
	&
	x^{-n_{2,\tilde 3,\tilde 5}-1} \bar x^{b-n_{1,\tilde 4,\tilde 5}-1} \Gamma\left[P_2\right] \mathsf{f}^{0:3;3;2}_{2:1;0;0} \left( \left. Q_2 \right\rvert 1, 1, \bar x \right)
\label{eq:MM-xb}
\end{align}
and
\begin{align}
	G(\{n_i\};\underline{a},\underline{b};D)
	=
	\Gamma\left[P'_1\right] \mathsf{f}^{0:3;3;3}_{2:1;0;1} \left(  \left. Q_1' \right\rvert 1, 1, 1 \right)
	+\Gamma\left[P'_2\right] \mathsf{f}^{0:3;3;3}_{2:1;0;1} \left( \left. Q_2' \right\rvert 1, 1, 1 \right).
\label{eq:MM_ab}
\end{align}
Here, $\mathsf{f}^{p_0:p_1,p_2,p_3}_{q_0:q_1,q_2,q_3}(Q\vert z_1,z_2,z_3)$ is a generalized Lauricella function defined in appendix~\ref{app:HGFs-defs}, eq.~\eqref{eq:gen-Lauricella-def}. The arrays of the parameters of the two-row $\Gamma$ functions, $P_i=P_i(\{n_i\},b,D)$, $P'_i=P'_i(\{n_i\}, \allowbreak a,\allowbreak b, \allowbreak D)$, and Lauricella functions $Q_i=Q_i(\{n_i\},b,D)$, $Q'_i=Q'_i(\{n_i\},a,b,D)$, $i=1,2$ are given explicitly in eqs.~\eqref{eq:PQ}--\eqref{eq:P'Q'}.

At least in some cases the threefold hypergeometric series in eqs.~\eqref{eq:MM-xb} and \eqref{eq:MM_ab} reduce to simpler KdF functions. To see this, we can apply the method of ref.~\cite{Bierenbaum2003} to the original two-loop integral \eqref{eq:2loopscint} (see also refs.~\cite{belokurov1984calculating6082,Usyukina1989} with less general results). Firstly, we write the Mellin--Barnes representation for one of the subgraphs of the two-loop diagram:
\begin{gather}
	\left. \int \frac{\mathrm{d}^D k_1 \, \delta\left(x-\tilde nk_1\right)  }{ \mathscr{D}_1^{n_1} \mathscr{D}_3^{n_3} \mathscr{D}_5^{n_5}} \right|_{\tilde nk_2 = y}
\notag\\{}=
	i \pi^{D/2} (-)^{n_{1,3,5}} (-p^2)^{-n_{1,3,\tilde 5}} \int \frac{\mathrm{d}q_1}{2\pi i} \int \frac{\mathrm{d}q_2}{2\pi i}\; A^{q_1} B^{q_2}
				\Gamma\begin{bmatrix} n_{1,3,\tilde 5} + q_{1,2},\,-q_1,\, -q_2 \\ n_1,\,n_3,\,n_5 \end{bmatrix}
\notag\\
	{}\times \int_0^1 \mathrm{d}\beta_1 \int_0^1 \mathrm{d}\beta_3 \int_0^1 \mathrm{d}\beta_5\,
	\beta_1^{-n_{3,\tilde 5}-q_1-1} \beta_3^{-n_{1,\tilde 5}-q_2-1} \beta_5^{n_5+q_{1,2}-1}
\notag\\{}\times
	\delta(1-\beta_1-\beta_3-\beta_5) \delta(x-\beta_1-y \beta_5),
\label{eq:triangle-MB}
\end{gather}
where $A=\mathscr{D}_4/p^2$, $B=\mathscr{D}_2/p^2$ are dimensionless parameters and the denominators $\mathscr{D}_i$, $i=1,...,5$ were defined earlier in eq.~\eqref{eq:denoms}. Note that integrating eq.~\eqref{eq:triangle-MB} over $x$ leads to the Mellin--Barnes representation for the triangle diagram given in refs.~\cite{Boos:1987bg,Boos:1990rg,Usyukina1975} since the Dirac delta becomes identity and
\begin{gather}
	\int_0^1 \mathrm{d}\beta_1 \int_0^1 \mathrm{d}\beta_3 \int_0^1 \mathrm{d}\beta_5\,
	\beta_1^{-n_{3,\tilde 5}-q_1-1} \beta_3^{-n_{1,\tilde 5}-q_2-1} \beta_5^{n_5+q_{1,2}-1}
	\delta(1-\beta_1-\beta_3-\beta_5)
\notag\\
	{}=
	\Gamma\begin{bmatrix}-n_{3,\tilde 5}-q_1,\,-n_{1,\tilde 5}-q_2,\, n_5+q_{1,2} \\ -n_{1,\tilde 3,\tilde 5} \end{bmatrix}.
\label{eq:beta-int}
\end{gather}

Secondly, we insert eq.~\eqref{eq:triangle-MB} into the original two-loop integral \eqref{eq:2loopscint} to get the Mellin--Barnes integral for the function $G(\{n_i\};x,y;D)$:
\begin{gather}
	G(\{n_i\};x,y;D) = \int \frac{\mathrm{d}q_1}{2\pi i} \int \frac{\mathrm{d}q_2}{2\pi i}\;
	\Gamma\begin{bmatrix}-q_1,\, -q_2,\, n_{1,3,\tilde 5}+q_{1,2},\, n_{2,\tilde 4}-q_{1,2} \\
									n_1,\,n_2-q_2,\,n_3,\,n_4-q_1,\,n_5
				\end{bmatrix}
	y^{q_1-n_{\tilde 4}-1} \bar{y}^{q_2-n_{\tilde 2}-1}
\notag\\
	{}\times
	\int_0^1 \mathrm{d}\beta_1 \int_0^1 \mathrm{d}\beta_3 \int_0^1 \mathrm{d}\beta_5\,
	\beta_1^{-n_{3,\tilde 5}-q_1-1} \beta_3^{-n_{1,\tilde 5}-q_2-1} \beta_5^{n_5+q_{1,2}-1}
\notag\\
	{}\times	\delta(1-\beta_1-\beta_3-\beta_5) \delta(x-\beta_1-y \beta_5).
\label{eq:G-MB}
\end{gather}
For the zeroth moment $G(\{n_i\};\underline{0},y;D)$ the last Dirac delta in the above expression drops out and with the help of eq.~\eqref{eq:beta-int} we arrive at a normal looking twofold Mellin--Barnes integral that can be evaluated in terms of double series:
\begin{align}\notag
	G(\{n_i\};\underline{0},y;D)
	=\Gamma\left[\Xi_0\right] \Biggl\{ {}& \sum_{k=1}^6 (-)^k y^{\alpha_k-1} \bar y^{\beta_k-1} \Gamma\left[\Xi_k\right] \mathsf{f}^{2:2;1}_{1:2;1} \left( \Phi_k \left\lvert \, \bar y, y \right. \right)
\notag\\
	&{}+\sum_{k=7}^8 (-)^k y^{\alpha_k-1} \bar y^{\beta_k-1} \Gamma\left[\Xi_k\right] \mathsf{f}^{1:2;3}_{2:0;1} \left( \Phi_k \left\lvert \, y, -\frac{y}{\bar y} \right. \right)
\notag\\
	&{}-\sum_{k=9}^{10} (-)^k y^{\alpha_k-1} \bar y^{\beta_k-1} \Gamma\left[\Xi_k\right] {}_3f_2 \left( \Phi_k \left\lvert \, y \right. \right) \Biggr\}.
\label{eq:g0y}
\end{align}
The same is true for twofold moments:
\begin{align}
	G(\{n_i\};\underline{0},\underline{b};D)
	= \Gamma\left[\Xi_0'\right] \left\{ \sum_{k=1}^8 (-)^k \Gamma\left[\Xi_k'\right] \mathsf{f}^{2:3;2}_{2:2;1} \left( \left. \Phi_k' \right\rvert 1, 1 \, \right)
	-\sum_{k=9}^{10} (-)^k \Gamma\left[\Xi_k'\right] {}_4f_3 \left( \left. \Phi_k' \right\rvert 1 \,\right) \right\}.
\label{eq:g0b}
\end{align}
In eqs.~\eqref{eq:g0y} and \eqref{eq:g0b} the parameters $\alpha_k = \alpha_k (\{n_i\},D)$, $\beta_k = \beta_k (\{n_i\},D)$, $\Xi_k = \Xi_k (\{n_i\},D)$, $\Xi_k = \Xi_k (\{n_i\},b,D)$, $\Phi_k = \Phi_k(\{n_i\},D)$, and $\Phi_k' = \Phi_k'(\{n_i\},b,D)$, $k=1,\dots,10$ are listed in eqs.~\eqref{app:ablist}--\eqref{app:Phi'list} of appendix \ref{App:C}. By virtue of the symmetries \eqref{eq:2loopsym1}, it is easy to see that a similar reduction is possible for moments $G(\{n_i\};x,\underline{0};D)$ and $G(\{n_i\};\underline{a},\underline{0}; \allowbreak D)$. Note also that the Mellin--Barnes integral \eqref{eq:G-MB} allows us also to express in terms of KdF functions the higher moments $G(\{n_i\};x,\underline{N};D)$, $G(\{n_i\};\underline{N},y;D)$, $G(\{n_i\};\underline{a},\underline{N}; \allowbreak D)$, and $G(\{n_i\};\underline{N},\underline{b}; \allowbreak D)$, where $N$ is a natural number. In the special case of $b=0$, eqs.~\eqref{eq:g0b} and \eqref{app:Xi'list}--\eqref{app:Phi'list} give an expression for $G(\{n_i\};\allowbreak \underline{0},\underline{0};\allowbreak D)$ that coincides after some rearrangement with the result obtained in \cite{Bierenbaum2003}.

%=====================================================================================
%=====================================================================================
%=====================================================================================

\section{\label{sec:special}Important special case of indices \boldmath$n_{i}=1$, at $i=1,\ldots 4$ and any $n_5=n$}

This set of indices appears, for example, in the QCD calculation of a leading 3-loop contribution to the two--quark-current correlator of order $O(\beta_0 \alpha_s^2)$.

\begin{figure}[h]
\centering
\includegraphics{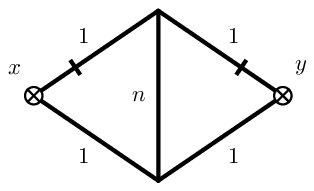}
\caption{\label{fig:1111n}Kite diagram with four vertices of external lines being equal to 1.}
\end{figure}

%=====================================================================================
%=====================================================================================
%=====================================================================================

\subsection{Correlator $G(1,1,1,1,n;x,y;D)$}

If all indices of the external edges of the diagram are equal to one (see figure \ref{fig:1111n}), the corresponding integral splits into two univariate functions---a hypergeometric function of the conformal ratio $z$ and a power of the difference $x-y$:
\begin{gather}\label{eq:1111n-fact}
	G(1,1,1,1,n;x,y;D)
=
	\frac{\hat{\mathbf{S}} f(n;z;D)}{\lvert x-y \rvert^{-\omega/2}},
\end{gather}
\begin{align}
	f(n;z;D) = {}&\Gamma\begin{bmatrix}2+\dot{n}-\lambda,\; \dot{n},\; 1-\dot{n} \\ n,\; \lambda \end{bmatrix}
		z^{\lambda-1} \bar z^{2-\lambda}
\notag\\ \label{eq:f(z;D)nu}
&{} \times
	\Biggl[ {}_3f_2 \left( \left. \begin{matrix} 1, \; 1, \; \lambda \; \\ 1-\dot{n}, \; \dot{n}+2 \;\end{matrix} \right| \bar z \right)
	-\bar z^{\dot{n}} {}_2f_1 \left( \left. \begin{matrix} n, \; \dot{n}+1 \; \\ 2(\dot{n}+1) \; \end{matrix} \right| \bar z \right)
	\Biggr] \Theta(\bar z)  ,
\end{align}
where $z=(\bar x y)/(x \bar y)$ is the conformal ratio defined earlier, $\omega = 2(D-4-n)$ is degree of divergence of the integral, $\lambda = D/2-1$, $\dot{n} = n-\lambda$, and $\hat{\mathbf{S}} T(x_1,x_2)=T(x_1,x_2)+T(\bar x_1, \bar x_2)$, i.e. $\hat{\mathbf{S}} f(z)=f(z)+f(1/z)$. In eq.~\eqref{eq:f(z;D)nu} we introduce renormalized hypergeometric functions for the sake of brevity (see appendix \ref{app:HGFs-defs}):
\begin{align}
	{}_pf_q \left( \left. \begin{matrix} \mathbf a \\ \mathbf b \end{matrix}\; \right| z \right) = \Gamma \begin{bmatrix} \mathbf a \\ \mathbf b \end{bmatrix} {}_pF_q \left( \left. \begin{matrix} \mathbf a \\ \mathbf b \end{matrix}\; \right| z \right),
\end{align}
where $\mathbf a = a_1,\,\dots, a_p$,  $\mathbf b = b_1,\,\dots, b_q$ are parameters of the function.

For $n=1$, eqs.~\eqref{eq:1111n-fact} and (\ref{eq:f(z;D)nu}) reduce to
\begin{gather}
		G(1,1,1,1,1;x,y;D)
=
	\frac{\hat{\mathbf{S}} f(1;z;D)}{\lvert x-y \rvert^{5-D}},
\\
	f(1,z;D) = \Gamma\left(3-2\lambda\right) \Gamma\left(\lambda-1\right) \Gamma\left(1-\lambda\right)
		\Theta(\bar z) \left[ I_{\bar z}\left(2-\lambda,\lambda-1\right) - I_{\bar z}\left(3-2\lambda,\lambda-1\right) \right],
\end{gather}
where $I_{\bar z}(a,b)$ is an incomplete Euler B function normalized by the complete one:
\begin{gather}
	I_x(a,b) = \frac{\text{B}_x(a,b)}{\text{B}_1(a,b)}.
\end{gather}

%=============================================================================
%=============================================================================
%=============================================================================

\subsection{\label{sec:1111n-moments}Mellin moments of the correlator as the Kamp\'{e} de F\'{e}riet functions}

Both one- and twofold moments of the function \eqref{eq:1111n-fact} can be written in terms of the KdF functions for arbitrary real orders $\underline a$ and $\underline b$ of the moments:

\begin{align}
	G(1,1,1,1&,n;x,\underline b;D)
=
x^b (x \bar x)^{\lambda-\dot{n} -1} \Gamma\begin{bmatrix} \dot{n},\; 1-\dot{n},\; 2+\dot{n} -\lambda \\
	                                                                                           n
	                                                              \end{bmatrix}
\notag\\
	&{} \times \Biggl\{ -\Gamma\begin{bmatrix} b+\lambda \\
	                                                                                           \lambda,\; b+\lambda-\dot{n}
	                                                                    \end{bmatrix}
	\mathsf f^{1:1;2}_{1:0;1} \left(  \left.\begin{matrix}1&: & b+\lambda-\dot{n}&;& n,\,\dot{n}+1 \, \\
																			 1+\lambda+b&:&\text{---}&;&2(\dot{n}+1) \,
														\end{matrix} \right\rvert x, 1 \right)
\notag\\
	&\hphantom{{} \times \Biggl\{}{}-\frac{1}{\Gamma(b+\lambda-\dot{n})} \mathsf f^{1:1;2}_{1:0;1} \left(  \left.\begin{matrix}1&: & b+\lambda-\dot{n}&;& n,\,\dot{n}+1 \, \\
																			 1+\lambda&:&\text{---}&;&2(\dot{n}+1) \,
														\end{matrix} \right\rvert \bar x, 1 \right)
\notag\\
	&\hphantom{{} \times \Biggl\{}{}+\frac{\bar x^{\lambda+b}}{\Gamma(\lambda)}
	\mathsf f^{1:1;2}_{1:0;1} \left(  \left.\begin{matrix}1&: & b+\lambda&;& 1,\,\lambda \, \\
																			 1+\lambda-\dot{n}+b&:&\text{---}&;&\dot{n}+2 \,
														\end{matrix} \right\rvert x, 1 \right)
\notag\\
	&\hphantom{{} \times \Biggl\{}{} + \frac{1}{\Gamma(b+\lambda-\dot{n})}
	\mathsf f^{1:1;3}_{1:0;2} \left(  \left.\begin{matrix}1-\dot{n}&: & b+\lambda-\dot{n}&;& 1,\,1,\,\lambda \, \\
																			 1+\lambda-\dot{n}&:&\text{---}&;&1-\dot{n},\,\dot{n}+2 \,
														\end{matrix} \right\rvert \bar x, 1 \right)
	\Biggr\},
\label{eq:1111n-xb}
\end{align}

\begin{align}
	G(&1,1,1,1,n;\underline a,\underline b;D)
=
\Gamma\begin{bmatrix}
					\dot{n},\; 1-\dot{n},\; 2+\dot{n} -\lambda \\
	                 n
	        \end{bmatrix}
\notag\\
	&{} \times \Biggl\{ -\Gamma\begin{bmatrix} b+\lambda,\; \lambda-\dot{n} \\
	                                                               \lambda,\; b+\lambda-\dot{n}
	                                        \end{bmatrix}
	\mathsf f^{1:2;2}_{1:1;1} \left(  \left.\begin{matrix}1&: & b+\lambda-\dot{n},\, a+b+\lambda-\dot{n}&;& n,\,\dot{n}+1 \, \\
																			 1+\lambda+b&:&a+b+2\lambda-2\dot{n}&;&2(\dot{n}+1) \,
														\end{matrix} \right\rvert 1, 1 \right)
\notag\\
	&\hphantom{{} \times \Biggl\{}{} {}-\frac{\Gamma(a+b+\lambda-\dot{n})}{\Gamma(b+\lambda-\dot{n})} \mathsf f^{1:2;2}_{1:1;1} \left(  \left.\begin{matrix}1&: & b+\lambda-\dot{n}, \lambda-\dot{n}&;& n,\,\dot{n}+1 \, \\
																			 1+\lambda&:&a+b+2\lambda-2\dot{n}&;&2(\dot{n}+1) \,
														\end{matrix} \right\rvert 1, 1 \right)
\notag\\
	&\hphantom{{} \times \Biggl\{}{} {}+\frac{\Gamma(b+2\lambda-\dot{n})}{\Gamma(\lambda)}
	\mathsf f^{1:2;2}_{1:1;1} \left(  \left.\begin{matrix}1&: & b+\lambda,\, a+b+\lambda-\dot{n}&;& 1,\,\lambda \, \\
																			 1+\lambda-\dot{n}+b&:& a+2b+3\lambda-2\dot{n} &;&\dot{n}+2 \,
														\end{matrix} \right\rvert 1, 1 \right)
\notag\\
	&\hphantom{{} \times \Biggl\{}{} {} + \frac{\Gamma(a+b+\lambda-\dot{n})}{\Gamma(b+\lambda-\dot{n})}
	\mathsf f^{1:2;3}_{1:1;2} \left(  \left.\begin{matrix}1-\dot{n}&: & b+\lambda-\dot{n},\, \lambda-\dot{n}&;& 1,\,1,\,\lambda \, \\
																			 1+\lambda-\dot{n}&:& a+b+2\lambda-2\dot{n} &;&1-\dot{n},\,\dot{n}+2 \,
														\end{matrix} \right\rvert 1, 1 \right)
	\Biggr\}.
\label{eq:1111n-ab}
\end{align}

%=============================================================================
%=============================================================================
%=============================================================================

\subsection{Reduction to hypergeometric series in one variable}

At least some of KdF functions in eqs.~\eqref{eq:1111n-xb} and \eqref{eq:1111n-ab} reduce to onefold hypergeometric series:
\begin{align}
\Gamma\begin{bmatrix}  2+\dot{n} -\lambda,\; b+\lambda \\
	                                 n,\; -b
	         \end{bmatrix}
	&\mathsf f^{1:1;2}_{1:0;1} \left(  \left.\begin{matrix}1&: & b+\lambda-\dot{n}&;& n,\,\dot{n}+1 \, \\
																			 1+\lambda+b&:&\text{---}&;&2(\dot{n}+1) \,
														\end{matrix} \right\rvert x, 1 \right)
\notag\\
{}=\frac{\pi}{\sin\left[\pi (\lambda-\dot{n}) \right]}
\Biggl\{
&\frac{\pi \sin\left[\pi (b+\dot{n}) \right]}{\sin(\pi b) \sin\left[\pi (b-\dot{n}) \right] (\dot{n} -\lambda)} \frac{1}{\Gamma(n)\Gamma(-b)} \bar x^{\dot{n}}
 {}_2f_1  \left(  \left.\begin{matrix}
								-\dot{n},\, b+\lambda \, \\
								1-\dot{n}+b \,
							\end{matrix} \right\rvert x \right)
\notag\\
&{}-\frac{\pi \cos(\pi \dot{n})}{\sin\left[\pi (b-\dot{n}) \right] (\dot{n} -\lambda)} \frac{1}{\Gamma(n)\Gamma(-b)} x^{-b} (x \bar x)^{\dot{n}} 
 {}_2f_1  \left(  \left.\begin{matrix}
								n,\, -b \, \\
								1+\dot{n}-b \,
							\end{matrix} \right\rvert x \right)
\notag\\ 
&{}-2 \cos(\pi \dot{n})
 {}_3f_2  \left(  \left.\begin{matrix}
								1,\, -2\dot{n},\, b+\lambda-\dot{n} \, \\
								1-\dot{n},\, \lambda-\dot{n} \,
							\end{matrix} \right\rvert \bar x \right)
\Biggr\},
\label{eq:f112101-red-1}
\end{align}
\begin{align}
\mathsf f^{1:1;2}_{1:0;1} &\left(  \left.\begin{matrix}1&: & b+\lambda-\dot{n}&;& n,\,\dot{n}+1 \\
																			 1+\lambda&:&\text{---}&;&2(\dot{n}+1)
														\end{matrix} \right\rvert \bar x, 1 \right)
\notag\\
={}&\Gamma\begin{bmatrix}\dot{n}+1,\; -\dot{n},\; \lambda-\dot{n}-1,\; n \\
								-2 \dot{n}-1,\; 2 ( \dot{n}+1),\; \lambda
			\end{bmatrix}
{}_3f_2 \left(  \left.\begin{matrix}1,\; b+\lambda-\dot{n},\; -2\dot{n}\, \\
											1-\dot{n},\; \lambda-\dot{n}\,
								\end{matrix} \right\rvert \bar{x} \right)
\notag\\
&{}+x^{\dot{n}-\lambda-b} \Gamma\begin{bmatrix}\lambda-\dot{n}-1,\; -\dot{n} \\
																	\lambda
											\end{bmatrix}
{}_3f_2 \left(  \left.\begin{matrix}1,\; \lambda,\; b+\lambda-\dot{n}\, \\
											1-\dot{n},\; \lambda-\dot{n}\,
								\end{matrix} \right\rvert -\frac{\bar{x}}{x} \right),
\label{eq:f112101-red-2}
\end{align}
\begin{align}
&\mspace{-100mu}-\Gamma\begin{bmatrix} \dot{n},\; 1-\dot{n},\; 2+\dot{n} -\lambda,\; b+\lambda,\; \lambda-\dot{n} \\
	                                 n,\; \lambda,\; b+\lambda-\dot{n}
	         \end{bmatrix}
\notag\\
{}\times
	\mathsf f^{1:2;2}_{1:1;1} &\left(  \left.\begin{matrix}1&: & b+\lambda-\dot{n},\, a+b+\lambda-\dot{n}&;& n,\,\dot{n}+1 \, \\
																			 1+\lambda+b&:&a+b+2\lambda-2\dot{n}&;&2(\dot{n}+1) \,
														\end{matrix} \right\rvert 1, 1 \right)
\notag\\
&\mspace{-60mu}{}={}\frac{\pi^2}{\sin\left(\pi\dot{n}\right) \sin\left[\pi (\lambda-\dot{n}) \right]}
\notag\\
{}\times
\Biggl\{
&\frac{\pi \sin\left[\pi (b+\dot{n}) \right]}{\sin(\pi b) \sin\left[\pi (\dot{n}-b) \right] (\dot{n} -\lambda)} \frac{1}{\Gamma(n)\Gamma(-b)}
 {}_3f_2  \left(  \left.\begin{matrix}
								-\dot{n},\, b+\lambda,\, a+b+\lambda-\dot{n} \, \\
								1-\dot{n}+b,\, a+b+2\lambda-\dot{n} \,
							\end{matrix} \right\rvert 1 \right)
\notag\\
&{}+2 \cos(\pi \dot{n}) \frac{\Gamma(a+b+\lambda-\dot{n})}{\Gamma(\lambda)}
 {}_3f_2  \left(  \left.\begin{matrix}
								1,\, -2\dot{n},\, b+\lambda-\dot{n} \, \\
								1-\dot{n},\, a+b+2\lambda-2\dot{n} \,
							\end{matrix} \right\rvert 1 \right)
\notag\\
&{}+\frac{\pi \cos(\pi \dot{n})}{\sin\left[\pi (b-\dot{n}) \right] (\dot{n} -\lambda)} \frac{1}{\Gamma(n)\Gamma(-b)}
 {}_3f_2  \left(  \left.\begin{matrix}
								n,\, -b,\, a+\lambda \, \\
								1+\dot{n}-b,\, a+2\lambda \,
							\end{matrix} \right\rvert 1 \right)
\Biggr\}.
\end{align}
The proofs of the above reductions are reserved for the appendix \ref{app:110f121}. All other KdF functions in eqs.~\eqref{eq:1111n-xb} and \eqref{eq:1111n-ab} reduce to generalized hypergeometric functions for $b=0$:
\begin{align}
	G(1,1,1,1,n;x,\underline 0;D)
={}&
	{}-\Gamma\begin{bmatrix}-\dot{n},\; 1+\dot{n}-\lambda \\ n \end{bmatrix} (x \bar x)^{\lambda-1}
\notag\\
	&{} \times
	\Biggl\{
		\Gamma\begin{bmatrix}n,\; \dot{n},\; \dot{n} ,\; 1-\dot{n} ,\; 1-\dot{n} \\ \lambda,\; 2 \dot{n},\; 1-2 \dot{n} \end{bmatrix}
		+\hat{\mathbf{S}} \Biggl[ x^{-\dot{n}}
		{}_3f_2 \left( \left. \begin{matrix} 1,\; \lambda, \; -\dot{n} \\ 1-\dot{n},\; \lambda-\dot{n} \end{matrix} \right\rvert x \right) \Biggr]
	\Biggr\}.
\label{eq:1111n-x0}
\end{align}
An immediate consequence of the variable splitting in eq.~\eqref{eq:1111n-fact} is that a moment $(x,\underline{N+b})$ shifted by a natural number $N$ with respect to a lower moment $(x,\underline{b})$ can be obtained by differentiating the latter one:
\begin{multline}\label{eq:1111n-mom-shift}
	G(1,1,1,1,n;x,\underline{N+b};D)
\\
{}=
	(x \bar x)^{1+\omega/2} x^{b+N} \sum_{k=0}^N \begin{pmatrix} N \\ k \end{pmatrix} \frac{\bar x^k}{(2+\omega/2+b)_k}
	\frac{\partial^k}{\partial \bar x^k} \biggl[ (x \bar x)^{-1-\omega/2} x^{-b} G(1,1,1,1,n;x,\underline{b};D) \biggr].
\end{multline}
This implies that any moment $G(1,1,1,1,n;x,\underline N;D)$ for a natural $N$ is a finite sum of functions $\vphantom{F}_3F_2(x)$, $\vphantom{F}_3F_2(\bar x)$, and simpler functions.

Setting $N$ to 1 and integrating $\eqref{eq:1111n-mom-shift}$ with $x^a$ by parts, we obtain a simple recurrence relation for twofold moments:
\begin{multline}\label{eq:1111n-RR}
	(\omega/2+a+b+2) G(1,1,1,1,n;\underline{a},\underline{b};D)
\\
	{} = (\omega/2+a+2) G(1,1,1,1,n;\underline{a+1},\underline{b};D)
	+ (\omega/2+b+2) G(1,1,1,1,n;\underline{a},\underline{b+1};D).
\end{multline}
Deriving the above relation, we assumed that the limits of $x^{1+a} \bar x G(1,1,1,1,n;x,\underline{b};D)$ at the endpoints are zero. The recurrence relation allows us to express all moments $(\underline{a+k}, \underline{b+l})$ for any natural numbers $k$ and $l$ through a set of independent moments chosen, for instance, as $(\underline{a+2k},\underline b)$, $k=1$, 2, \dots.

Evaluating $\underline a$-moments of eq.~\eqref{eq:1111n-x0}, we arrive at
\begin{align}
	G(1,1,1,1,n;\underline a,\underline 0;D)
={}&
	-\Gamma\left(-\dot{n}\right) \Gamma\left(\dot{n}\right) \Gamma\left(1+\dot{n}-\lambda\right)
\notag\\
	&{} \times
	\Biggl\{ \Gamma\begin{bmatrix}\lambda\\ n,\; \dot{n} \end{bmatrix}
	{}_4f_3 \left( \left. \begin{matrix} 1,\; \lambda, \; -\dot{n}, \; a+\lambda-\dot{n} \, \\ 1-\dot{n},\; \lambda-\dot{n}, \; a+2\lambda-\dot{n} \,\end{matrix} \right\rvert 1 \right)
\notag\\
&\hphantom{{}\times\Biggl\{}+ \Gamma\begin{bmatrix}a+\lambda\\ n,\; \dot{n} \, \end{bmatrix}
	{}_3f_2 \left( \left. \begin{matrix} 1,\; \lambda, \; -\dot{n} \, \\ 1-\dot{n}, \; a+2\lambda-\dot{n} \, \end{matrix} \right\rvert 1 \right)
\notag\\\label{eq:1111n-a0}
	& \hphantom{{}\times\Biggl\{} + \Gamma\begin{bmatrix}\dot{n},\; 1-\dot{n}, \; 1-\dot{n}, \;a+\lambda \\  1-2\dot{n},\; 2 \dot{n},\; a+2\lambda \end{bmatrix}
	\Biggr\}.
\end{align}
This and the recurrence relation \eqref{eq:1111n-RR} give also all twofold moments $G(1,1,1,1,n;\underline a,\underline N;D)$ for any natural $N$. The moments $G(1,1,1,1,n;\underline a,\underline N;D)$ can always be represented as a sum of $_4f_3(1)$ and simpler functions.

Finally, to compare eq.~\eqref{eq:1111n-a0} with earlier results in the literature, we can write the following expression for the special case $a=b=0$ that is valid for arbitrary $D$ and $n$:
\begin{gather}
	G(1,1,1,1,n;\underline 0,\underline 0;D)
=
	-2\Gamma\left(-\dot{n}\right) \Gamma\left(1-\lambda+\dot{n}\right) \Gamma\left(\lambda\right)
\notag\\
	{} \times
	\Biggl\{ \Theta(2\lambda-n-1) \Biggl(
	\frac{1}{\Gamma(n)} {}_3f_2 \left( \left. \begin{matrix} 1,\; \lambda, \; -\dot{n} \, \\ 1-\dot{n},\; 2\lambda-\dot{n} \, \end{matrix} \right\rvert 1 \right) + \frac{\pi}{\Gamma(2\lambda)} \cot\left( \pi \dot{n} \right) \Biggr)
\notag\\
	{} +
	\Theta(1-2\lambda+n) \Biggl( \Gamma\begin{bmatrix}\lambda \\ 2\lambda,\; 2\lambda-\dot{n}-1 \end{bmatrix}
	{}_3f_2 \left( \left. \begin{matrix} 1,\; 2\lambda, \; 1+\dot{n} \, \\ 1+n,\; 2+\dot{n} \, \end{matrix} \right\rvert 1 \right) + \frac{\pi}{\Gamma(2\lambda)} \cot\left[ \pi (\lambda-\dot{n} ) \right]
	 \Biggr) \Biggr\}.
\label{eq:1111n-00}
\end{gather}
The above equation reproduces the already known results \cite{Kazakov1985, Broadhurst:1996ur, 1996PhLB..375..240K} in its different domain of applicability. Here, we have taken into account the fact that the function $_3f_2(1)$ converges if and only if its parametric excess---the difference between the sums of all lower and all upper parameters---is positive. The Heaviside step functions ``switch off'' the corresponding $_3f_2(1)$, if it diverges, which guarantees that the above representation is valid for all $\lambda$ and $n$. Note that the factors accompanying the step functions are identically equal in the band $D-3<n<D-2$. The equality can be proven by sequentially applying the identities \eqref{fwc:07.27.17.0037.01} and \eqref{fwc:07.27.17.0038.01} to the first ${}_3f_2(1)$ in eq.~\eqref{eq:1111n-00}. Note also that the factor accompanying the step function of $1-2\lambda+n$ is exactly the expression found by Kotikov \cite{1996PhLB..375..240K}, while the other half of eq.~\eqref{eq:1111n-00} can be easily transformed to one of the representations suggested by Broadhurst et al.\ \cite{Broadhurst:1996ur} (see also eq.~(6.11) in ref.~\cite{2012IJMPA..2730018G}).\footnote{The function $_3f_2(1)$ multiplying $\Theta(2\lambda-n-1)$ differs from that in eq.~(6.11) of ref.~\cite{2012IJMPA..2730018G} in that two of its parameters are shifted by 1. We can always shift them back, since one of the upper parameters of $_3f_2(1)$ is equal to 1. To see this, one can simply write $_3f_2(1)$ as an integral of $_2f_1(1)$ and make use of the relations between consecutive $_2f_1(1)$ for the case with one of the parameters being equal to 1 (e.g.\ see eqs.~(2.9) and (2.10) in ref.~\cite{2006JHEP...04..056K}).}

%===================================================
%===================================================
%===================================================

\section{\label{sec:concl}Conclusion}

We evaluated the massless two-loop kite master integral $I(p;\{n_i\};x,y;D)$ (figure~\ref{fig:intro}). The integral as well as its Mellin moments naturally occur in calculating the correlator of two composite verticies. The integral is a function of indices $n_i$, $i=1,\dots,5$, space-time dimension $D$, and two Bjorken fractions $x,y \in [0,1]$. Compared with the ordinary two-loop master integral $I(p;\{n_i\};D)$,
the master integral considered here has two Dirac delta functions in the integrand [eq.~\eqref{eq:2loopscint}]. These factors restrict integration over loop four-momenta $k_i$, $i=1,2$ to the collinear subspaces with, e.g., $k_1^+ = x p^+$, $k_2^+ = y p^+$, and arbitrary $k_i^-$, $k_i^\perp$ in light-cone coordinates. On the other hand, the Dirac deltas give unities upon integrating $I(p;\{n_i\};x,y;D)$ over $x$ and $y$, and $I(p;\{n_i\};x,y;D)$ reduces to $I(p;\{n_i\};D)$. This double zeroth and other Mellin moments of $I(p;\{n_i\};x,y;D)$ cover all the Feynman integrals stemming from the kite diagrams in light-quark QCD---$I(p;\{n_i\};x,y;D)$ can be considered as a generating function in this sense.

Considering the integral $I(p;\{n_i\};x,y;D)$  in the $\alpha$ representation, we have evaluated the integral in terms of the hypergeometric Kamp\'{e} de F\'{e}riet (KdF) function of two variables, $\mathsf{f}(-z/\bar{z},1)$.
In some important cases with two natural indices (figure~\ref{fig:condition}), the KdF function reduces to a sum of univariate hypergeometric functions $_3F_2(\bar{z})$.
The hypergeometric part of the integral depends on only one combination of the Bjorken variables, the conformal ratio $z = (\bar{x}y)/(x\bar{y})$,
which could be a manifestation of the conformal symmetry.
We have also calculated one- and twofold Mellin moments $I(p;\{n_i\};x,y;D)$.
In the general case they are expressed through the generalized Lauricella functions.
For natural moments, however, the Lauricella functions reduce to simpler KdF functions,
which is proved most easily in their Mellin--Barnes representation.

We paid close attention to the special cases of $\{n_i\}=\{1,1,1,1,n\}$,
which appear in calculating two- and three-loop quark correlator
(more precisely, the part of the correlator proportional to $\beta_0$).
These master integrals and the Mellin moments thereof can always be expressed through the KdF functions reducing to $_4F_3$
and simpler functions at least in most of the practically important situations.
Taking the double zeroth Mellin moment of $I(p;1,1,1,1,n;x,y;D)$---a twofold integral over $x$ and $y$---yielded the well known expression for the two-loop integral $I(p;1,1,1,1,n;D)$ in terms of $_3F_2$. This can be viewed as a curious methodological byproduct of our consideration---evaluating directly hypergeometric Eulerian integrals occurring in the $\alpha$ representation provides us with the third alternative approach to derive the result for $I(p;1,1,1,1,n;D)$, the first two being:
solving functional equations coming from the star-triangle and integration-by-parts relations \cite{Kazakov1985,Broadhurst:1985vq,Broadhurst:1996ur},
and integrating with the help of expansion in the basis of Gegenbauer polynomials \cite{1996PhLB..375..240K}.
Also, we have derived the recurrence relations allowing us to express higher moments shifted by some natural numbers through a basis of lower moments. It should be noted that these recurrence relations in our approach are not a consequence of integration-by-parts relations as it is usually the case but of variable splitting in the integral, which is a product of function of the conformal ratio $z$ and function of the difference $x-y$.

%==================================================================
%==================================================================
%==================================================================

\acknowledgments

We would like to thank  A.~G. Grozin for reading the manuscript and valuable remarks and A.~V. Kotikov for the fruitful discussion. S.M. was supported in part by the BelRFFR--JINR, grant F18D-002. N.V. was supported by a grant from the Russian Scientific Foundation (project no. 18-12-00213).

%==================================================================
%==================================================================
%==================================================================

\appendix

%==================================================================
%==================================================================
%==================================================================

\section{\label{app:HGFs-defs}Hypergeometric functions---notation}

The most general hypergeometric series encountered in this work is the generalized Lauricella function $\mathsf{f}^{p_0:p_1,p_2,p_3}_{q_0:q_1,q_2,q_3}$ of three variables. In the convention we use, it is defined as a threefold series
\begin{multline}
\mathsf{f}^{p_0:p_1,p_2,p_3}_{q_0:q_1,q_2,q_3} 
\left(  \left.\begin{matrix}\mathbf{g}_0&: & \mathbf{a}_1&;& \mathbf{a}_2&;& \mathbf{a}_3\, \\
								\mathbf{h}_0&: & \mathbf{b}_1&;& \mathbf{b}_2&;& \mathbf{b}_3\,
			\end{matrix} \right\rvert z_1, z_2, z_3 \right)
\\{}=
\sum_{r_1,r_2,r_3\geqslant 0}
\Gamma\begin{bmatrix}\mathbf{A}_0, \mathbf{a}_1+r_1, \mathbf{a}_2+r_2, \mathbf{a}_3+r_3 \\
								\mathbf{B}_0, \mathbf{b}_1+r_1, \mathbf{b}_2+r_2, \mathbf{b}_3+r_3
			\end{bmatrix} \frac{z_1^{r_1}}{r_1!} \frac{z_2^{r_2}}{r_2!} \frac{z_3^{r_3}}{r_3!},
\label{eq:gen-Lauricella-def}
\end{multline}
where
\begin{gather*}
	\mathbf{a}_i = a_{i1}, \dots,  a_{ip_i},
	\qquad
	\mathbf{b}_i = b_{i1}, \dots,  b_{iq_i},
\\
	\mathbf{a}_i+r = a_{i1}+r, \dots,  a_{ip_i}+r,
	\qquad
	\mathbf{b}_i+r = b_{i1}+r, \dots,  b_{iq_i}+r \text{ for } i \geqslant 1,
\\
	\mathbf{g}_0= (a_{01}:\alpha_{1}^{1},\alpha_{1}^{2},\alpha_{1}^{3}), \dots, (a_{0p_0}:\alpha_{p_0}^{1},\alpha_{p_0}^{2},\alpha_{p_0}^{3}),
\\
	\mathbf{h}_0= (b_{01}:\beta_{1}^{1},\beta_{1}^{2},\beta_{1}^{3}), \dots,  (b_{0q_0}:\beta_{q_0}^{1},\beta_{q_0}^{2},\beta_{q_0}^{3}),
\\
	\mathbf{A}_0=a_{01} + \sum_{i=1}^3 \alpha_{1}^i r_i, \dots,  a_{0p_0} + \sum_{i=1}^3 \alpha_{p_0}^i r_i,
	\qquad
	\mathbf{B}_0=b_{01} + \sum_{i=1}^3 \beta_1^i r_i, \dots,  b_{0q_0} + \sum_{i=1}^3 \beta_{q_0}^i r_i.
\end{gather*}
The two-row gamma function is defined as follows:
\begin{gather}
	\Gamma\begin{bmatrix}a_1,\dots,\; a_p \\ b_1,\dots,\; b_q\end{bmatrix}
	= \frac{\prod_{i=1}^{p} \Gamma(a_i)}{\prod_{i=1}^{q} \Gamma(b_i)}.
\end{gather}

If one of its arguments vanishes, $z_i=0$, and all $\alpha_i^j$, $\beta_i^j$ are equal to 1, the Lauricella function reduces to a double series dubbed as the Kamp\'{e} de F\'{e}rriet (KdF) function:
\begin{gather}
\mathsf{f}^{p_0:p_1;p_2}_{q_0:q_1;q_2} \left(  \left.\begin{matrix}\mathbf{a}_0&: & \mathbf{a}_1&;& \mathbf{a}_2\, \\
																			 \mathbf{b}_0&: & \mathbf{b}_1&;& \mathbf{b}_2\,
														\end{matrix} \right\rvert z_1, z_2 \right)
=
\sum_{r_1,r_2 \geqslant 0}
\Gamma\begin{bmatrix}\mathbf{a}_0+r_1+r_2, \mathbf{a}_1+r_1, \mathbf{a}_2+r_2 \\
								\mathbf{b}_0+r_1+r_2, \mathbf{b}_1+r_1, \mathbf{b}_2+r_2
			\end{bmatrix} \frac{z_1^{r_1}}{r_1!} \frac{z_2^{r_2}}{r_2!},
\label{eq:gen-KdF-def}
\end{gather}
All Appell functions and univariate generalized hypergeometric functions belong to the class of KdF functions. In particular, the Appell function $F_1$ is
\begin{align}\label{eq:AppellF1}
	F_1 \left. \left( \begin{matrix}a;\;b_1,\; b_2\,\\ c \end{matrix} \right\rvert z_1, z_2 \right)
&{}=
\Gamma \begin{bmatrix}
			c  \, \\
			a, b_1, b_2 \,
			\end{bmatrix}
\mathsf{f}^{1:1;1}_{1:0;0} \left. \left( \begin{matrix}a&: & b_1&;& b_2 \, \\
																			 c&: & \text{---} &;& \text{---} \,
														\end{matrix} \right\rvert z_1, z_2 \right)
\notag\\
&{}=
	\sum_{r,s \geqslant 0} \frac{(a)_{r+s}}{(c)_{r+s}} \frac{(b_1)_r}{r!} \frac{(b_2)_s}{s!} z_1^r z_2^s,
\end{align}
where $(a)_r = \Gamma(a+r)/\Gamma(a)$ is the Pochhammer symbol. The generalized hypergeometric function is
\begin{gather}\label{eq:pfq-def}
	{}_p f_q \left( \left. \begin{matrix}\mathbf{a} \\ \mathbf{b} \end{matrix} \,\right\rvert z \right)
=
\mathsf{f}^{p:0;0}_{q:0;0} \left. \left( \begin{matrix} \mathbf{a}&: & \text{---}&;& \text{---} \, \\
																			 \mathbf{b}&: & \text{---} &;& \text{---} \,
														\end{matrix} \right\rvert z, 0 \right)
=
	\sum_{r \geqslant 0} \Gamma\begin{bmatrix} \mathbf{a}+r \\ \mathbf{b}+r \end{bmatrix} \frac{z^r}{r!},
\end{gather}

We often use the hypergeometric functions in the normalization of eqs.~\eqref{eq:gen-Lauricella-def}, \eqref{eq:gen-KdF-def}, and \eqref{eq:pfq-def}, because, with this convention, many expressions contain less gamma functions as factors accompanying hypergeometric functions. However, the hypergeometric functions are usually defined in the following normalization different from that we use in this paper \cite{srivastava1985multiple, exton1978handbook}:
\begin{align}
	{}_pF_q \left( \left. \begin{matrix} \mathbf a \\ \mathbf b \end{matrix}\; \right| z \right) = \Gamma \begin{bmatrix} \mathbf b \\ \mathbf a \end{bmatrix} {}_pf_q \left( \left. \begin{matrix} \mathbf a \\ \mathbf b \end{matrix}\; \right| z \right)
	=
	\sum_{r \geqslant 0} \frac{\prod_{i=1}^p (a_i)_r}{\prod_{i=1}^q (b_i)_r} \frac{z^r}{r!},
\end{align}
\begin{align}
\mathsf{F}^{p_0:p_1,p_2}_{q_0:q_1,q_2} \left(  \left.\begin{matrix}\mathbf{a}_0&: & \mathbf{a}_1&;& \mathbf{a}_2\,  \\
																			 \mathbf{b}_0&: & \mathbf{b}_1&;& \mathbf{b}_2\,
														\end{matrix} \right\rvert z_1, z_2 \right)
&{}=
\Gamma\begin{bmatrix}\mathbf{b}_0, \mathbf{b}_1, \mathbf{b}_2 \\
								\mathbf{a}_0, \mathbf{a}_1, \mathbf{a}_2
			\end{bmatrix}
\mathsf{f}^{p_0:p_1,p_2}_{q_0:q_1,q_2} \left(  \left.\begin{matrix}\mathbf{a}_0&: & \mathbf{a}_1&;& \mathbf{a}_2\,  \\
																			 \mathbf{b}_0&: & \mathbf{b}_1&;& \mathbf{b}_2\,
														\end{matrix} \right\rvert z_1, z_2 \right)
\notag\\
	&{}=
	\sum_{r_1, r_2 \geqslant 0} \left\{ 
	\frac{\prod_{i=1}^{p_0} (a_{0i})_{r_1+r_2}}{\prod_{i=1}^{q_0} (b_{0i})_{r_1+r_2}}
	\prod_{j=1}^2 \left[ \frac{\prod_{i=1}^{p_j} (a_{ji})_{r_j}}{\prod_{i=1}^{q_j} (b_{ji})_{r_j}} \frac{z_j^{r_j}}{r_j!} \right] \right\},
\end{align}
and
\begin{align}
\mathsf{F}^{p_0:p_1,p_2,p_3}_{q_0:q_1,q_2,q_3} 
&\left(  \left.\begin{matrix}\mathbf{g}_0&: & \mathbf{a}_1&;& \mathbf{a}_2&;& \mathbf{a}_3\, \\
								\mathbf{h}_0&: & \mathbf{b}_1&;& \mathbf{b}_2&;& \mathbf{b}_3\,
			\end{matrix} \right\rvert z_1, z_2, z_3 \right)
\notag\\
&=
\Gamma\begin{bmatrix}\mathbf{b}_0, \mathbf{b}_1, \mathbf{b}_2, \mathbf{b}_3 \\
								\mathbf{a}_0, \mathbf{a}_1, \mathbf{a}_2, \mathbf{a}_3
			\end{bmatrix}
\mathsf{f}^{p_0:p_1,p_2,p_3}_{q_0:q_1,q_2,q_3} 
\left(  \left.\begin{matrix}\mathbf{g}_0&: & \mathbf{a}_1&;& \mathbf{a}_2&;& \mathbf{a}_3\, \\
								\mathbf{h}_0&: & \mathbf{b}_1&;& \mathbf{b}_2&;& \mathbf{b}_3\,
			\end{matrix} \right\rvert z_1, z_2, z_3 \right)
\notag\\&=
\sum_{r_1,r_2,r_3\geqslant 0} \left\{ 
	\frac{\prod_{i=1}^{p_0} (a_{0i})_{N_i}}{\prod_{i=1}^{q_0} (b_{0i})_{M_i}}
	\prod_{j=1}^3 \left[ \frac{\prod_{i=1}^{p_j} (a_{ji})_{r_j}}{\prod_{i=1}^{q_j} (b_{ji})_{r_j}} \frac{z_j^{r_j}}{r_j!} \right] \right\},
\end{align}
where $N_i = \sum_{j=1}^3\alpha_i^j r_j$, $M_i = \sum_{j=1}^3\beta_i^j r_j$, and the multi-parameters $\mathbf{a}_i$, $\mathbf{b}_i$, $\mathbf{g}_0$, $\mathbf{h}_0$ are defined after eq.~\eqref{eq:gen-Lauricella-def}. Note that small- and capital-letter functions are different in their analytical properties with respect to the parameters. The function $_pF_q$ or $\mathsf{F}^{p_0:\dots}_{q_0:\dots}$ ($_p f_q$ or $\mathsf{f}^{p_0:\dots}_{q_0:\dots}$) does not exist when one of its lower (upper) parameters equals 0, $-1$, $-2, \dots $.

%==================================================================
%==================================================================
%==================================================================

Finally, for the sake of completeness of the paper, we list relations referred to in the text:
% 07.36.07.0001.01
\begin{align}
	F_1 \left. \left( \begin{matrix} a;b_1,b_2\,\\ c \end{matrix} \right\rvert z_1, z_2 \right)
	=
	\Gamma\begin{bmatrix} c \\ a,c-a \end{bmatrix}
	\int_0^1 \frac{ \mathrm{d}t\, t^{a-1} \bar{t}^{c-a-1}}{ \left(1-t z_1\right)^{b_1} \left(1-t z_2\right)^{b_2}},
	\quad
	\Re{c} > \Re{a} > 0 \text{ \cite{fwc:07.36.07.0001.01}};
\label{fwc:07.36.07.0001.01}
\end{align}

%http://functions.wolfram.com/07.36.17.0005.01
\begin{align}
	F_1 \left. \left( \begin{matrix} a;b_1,b_2\,\\ c \end{matrix} \right\rvert z_1, z_2 \right)
	=
	\bar{z}_1^{-b_1} \bar{z}_2^{c-a-b_2}
	F_1 \left. \left( \begin{matrix} c-a;b_1,c-b_1-b_2\,\\ c \end{matrix} \right\rvert \frac{z_2-z_1}{\bar{z}_1},z_2 \right)  \text{ \cite{fwc:07.36.17.0005.01}};
\label{fwc:07.36.17.0005.01}
\end{align}

%http://functions.wolfram.com/07.23.17.0061.01
\begin{align}
	_2F_1 \left. \left( \begin{matrix} a,b\,\\ c \end{matrix} \right\rvert z \right)
	={}&
	\Gamma\begin{bmatrix} c,a+b-c \\ a,b \end{bmatrix} \bar{z}^{c-a-b}
	 {}_2F_1\left. \left( \begin{matrix} c-a,c-b\,\\c-a-b+1 \end{matrix} \right\rvert \bar{z} \right)
	\notag\\
	&{}+ \Gamma\begin{bmatrix} c,c-a-b \\ c-a,c-b \end{bmatrix} {}_2F_1\left. \left( \begin{matrix} a,b\,\\a+b-c+1 \end{matrix} \right\rvert \bar{z} \right)  \text{ \cite{fwc:07.23.17.0061.01}};
\label{fwc:07.23.17.0061.01}
\end{align}

%http://functions.wolfram.com/07.31.07.0001.01
\begin{align}
	_{q+1}F_q \left. \left( \begin{matrix} a_1,\dots,a_{q+1}\,\\ b_1,\dots,b_q \end{matrix} \right\rvert z \right)
	={}&
	\Gamma\begin{bmatrix} b_q \\ a_{q+1},b_q-a_{q+1} \end{bmatrix}
	\notag\\ &{}\times
	\int_0^1 \mathrm{d}t\, t^{a_{q+1}-1} \bar{t}^{\;b_q-a_{q+1}-1} {}_qF_{q-1}\left. \left( \begin{matrix} a_1,\dots,a_q\,\\ b_1,\dots,b_{q-1} \end{matrix} \right\rvert tz \right),
	\notag\\
	& \Re b_q > \Re a_{q+1} > 0, \qquad \lvert \mathop{\mathrm{arg}} \bar{z} \rvert < \pi  \text{ \cite{fwc:07.31.07.0001.01}};
\label{fwc:07.31.07.0001.01}
\end{align}

%http://functions.wolfram.com/07.23.07.0001.01
\begin{gather}
	_2f_1 \left. \left( \begin{matrix} a,b \\ c \end{matrix} \right\rvert z \right)
	=
	\Gamma\begin{bmatrix} a \\ c-b \end{bmatrix}
	\int_0^1 \frac{\mathrm{d}t\, t^{b-1} \bar{t}^{\;c-b-1} }{( 1 - t z )^a},
	\qquad
	\Re c > \Re b > 0, \qquad \lvert \mathop{\mathrm{arg}} \bar{z} \rvert < \pi \text{ \cite{fwc:07.23.07.0001.01}};
\label{fwc:07.23.07.0001.01}
\end{gather}

%http://functions.wolfram.com/07.27.17.0037.01
\begin{gather}
\begin{aligned}
	_3f_2 \left. \left( \begin{matrix} a_1,a_2,a_3\,\\ b_1,b_2 \end{matrix} \right\rvert 1 \right)
	={}&
	\Gamma\begin{bmatrix} a_3,b_1-a_1-a_2,a_1+a_2-b_1+1\, \\ b_1-a_1,b_1-a_2,b_2-a_3\, \end{bmatrix}
	\Biggl\{ {}_3f_2 \left. \left( \begin{matrix} a_1,a_2,b_2-a_3\,\\a_1+a_2-b_1+1,b_2 \end{matrix} \right\rvert 1 \right)
\notag\\
	&{} - {}_3f_2 \left. \left( \begin{matrix} b_1-a_1,b_1-a_2,b_1+b_2-a_1-a_2-a_3\,\\ b_1-a_1-a_2+1,b_1+b_2-a_1-a_2 \end{matrix} \right\rvert 1 \right) \Biggr\},
\end{aligned}
\notag\\
	\Re\left(b_1+b_2-a_1-a_2-a_3\right)>0,
	\qquad
	\Re\left(a_3-b_1+1\right)>0 \text{ \cite{fwc:07.27.17.0037.01}};
\label{fwc:07.27.17.0037.01}
\end{gather}

%http://functions.wolfram.com/07.27.17.0039.01
\begin{gather}
\begin{aligned}
	_3f_2 \left. \left( \begin{matrix} a_1,a_2,a_3 \\ b_1,b_2 \end{matrix} \right\rvert 1 \right)
	={}&
	\Gamma\begin{bmatrix} a_2, a_3, a_2-b_1+1, a_3-b_1+1 \\ b_1, 1-b_1, b_2-a_1, a_1+a_2+a_3-b_1-b_2+1 \end{bmatrix}
\\&{}
	 \times {}_3f_2 \left. \left( \begin{matrix} a_1,a_1-b_1+1,a_1+a_2+a_3-b_1-b_2+1 \\ a_1+a_2-b_1+1,a_1+a_3-b_1+1 \end{matrix} \right\rvert 1 \right)
\\&{}
	+ {}_3f_2 \left. \left( \begin{matrix} a_1-b_1+1,a_2-b_1+1,a_3-b_1+1 \\ 2-b_1,-b_1+b_2+1 \end{matrix} \right\rvert 1 \right),
\end{aligned}
\notag\\
	\Re\left(b_1+b_2-a_1-a_2-a_3\right)>0,
	\qquad
	\Re\left(b_2-a_1\right)>0 \text{ \cite{fwc:07.27.17.0039.01}};
\label{fwc:07.27.17.0039.01}
\end{gather}

%http://functions.wolfram.com/07.27.17.0038.01
\begin{gather}
\begin{aligned}
	_3f_2 \left. \left( \begin{matrix} a_1,a_2,a_3 \\ b_1,b_2 \end{matrix} \right\rvert 1 \right)
	={}&
	\Gamma\begin{bmatrix} a_2, a_1-b_1+1, a_3-b_1+1 \\ b_2-a_2, b_1, 1-b_1 \\ \end{bmatrix}
	{}_3f_2 \left. \left( \begin{matrix} a_1,a_3,b_2-a_2 \\ a_1+a_3-b_1+1,b_2 \end{matrix} \right\rvert 1 \right)
\\
	&{}+ {}_3f_2 \left. \left( \begin{matrix} a_1-b_1+1,a_2-b_1+1,a_3-b_1+1 \\ 2-b_1,b_2-b_1+1 \end{matrix} \right\rvert 1 \right),
\end{aligned}
\notag\\
	\Re\left(b_1+b_2-a_1-a_2-a_3\right)>0,
	\qquad
	\Re\left(a_2-b_1+1\right)>0 \text{ \cite{fwc:07.27.17.0038.01}};
\label{fwc:07.27.17.0038.01}
\end{gather}

%http://functions.wolfram.com/07.23.03.0002.01
\begin{gather}
	{}_2f_1(a,b;c;1) = \Gamma\begin{bmatrix} a, b, c-a-b \\ c-a, c-b \end{bmatrix},
	\qquad
	\Re(c-a-b)>0 \text{ \cite{fwc:07.23.03.0002.01}};
\label{fwc:07.23.03.0002.01}
\end{gather}

%http://functions.wolfram.com/07.36.03.0006.01
\begin{gather}
	f_1 \left. \left( \begin{matrix} a;b_1,b_2\,\\ b_1+b_2 \end{matrix} \right\rvert z_1, z_2 \right)
	=
	\Gamma(b_2) \bar{z}_2^{\;-a} {}_2f_1 \left. \left( \begin{matrix} a, b_1 \\ b_1+b_2 \end{matrix} \right\rvert \frac{z_1-z_2}{\bar{z}_2} \right) \text{ \cite{fwc:07.36.03.0006.01}};
\label{fwc:07.36.03.0006.01}
\end{gather}

%http://functions.wolfram.com/07.36.03.0003.01
\begin{gather}
	f_1 \left. \left( \begin{matrix} a;b_1,b_2\,\\ c \end{matrix} \right\rvert z, 1 \right)
	=
	\Gamma\begin{bmatrix} c-b_2 \\ a \end{bmatrix}
	{}_2f_1 \left. \left( \begin{matrix} a, b_2 \\ c \end{matrix} \right\rvert 1 \right)
	{}_2f_1 \left. \left( \begin{matrix} a, b_1 \\ c-b_2 \end{matrix} \right\rvert z \right) \text{ \cite{fwc:07.36.03.0003.01}};
\label{fwc:07.36.03.0003.01}
\end{gather}

%http://functions.wolfram.com/07.27.17.0026.01
\begin{gather}
\begin{aligned}
	_3f_2 \left. \left( \begin{matrix} a_1,a_2,a_3 \\ b_1,b_2 \end{matrix} \right\rvert z \right)
	={}&
	 (-z)^{-a_1} \frac{\sin\left[\pi\left(b_1-a_1\right)\right] \sin\left[\pi\left(b_2-a_1\right)\right]}{\sin\left[\pi\left(a_2-a_1\right)\right] \sin\left[\pi\left(a_3-a_1\right)\right]}
\\{}&
	\times {}_3f_2 \left. \left( \begin{matrix} a_1, a_1-b_1+1, a_1-b_2+1 \\ a_1-a_2+1, a_1-a_3+1 \end{matrix} \right\rvert \frac{1}{z} \right)
\\{}&
	+ \text{ cyclic permutations of $\{a_1,a_2,a_3\}$},
\end{aligned}
\notag\\
	a_1-a_2 \notin \mathbb{Z},
	\qquad
	a_2-a_3 \notin \mathbb{Z},
	\qquad
	a_3-a_1 \notin \mathbb{Z},
	\qquad
	z \notin (0,1) \text{ \cite{fwc:07.27.17.0026.01}}.
\label{fwc:07.27.17.0026.01}
\end{gather}

%==================================================================
%==================================================================
%==================================================================

\section{\label{app:110f121}\boldmath KdF function $\mathsf{f}^{1:1;2}_{1:0;1}$}

The purpose of this appendix is to encapsulate some properties of the KdF function $\mathsf{f}_{1:1;0}^{1:2;1}(x,y)$ that has occurred in eqs.~\eqref{eq:FthKdF} and \eqref{eq:1111n-xb}. It admits three representations as the integrals of the Gauss and Appell functions, and $_3f_2$:
\begin{align}
	\mathsf{f}^{1:1;2}_{1:0;1} \left( \left.\begin{matrix}a&:&c&;&d,\,e\,\\ b&:& \text{---} &;&g\,\end{matrix} \right\rvert x, y \right)
	&{}= \Gamma\begin{bmatrix} c \\ b-a \end{bmatrix} \int_0^1 \mathrm{d}z\, \frac{ z^{a-1} \bar{z}^{b-a-1} }{ (1- x z)^c } {}_2f_1 \left( \left.\begin{matrix}d,\,e\,\\ g\,\end{matrix} \right\rvert yz \right)
\label{eq:110f121via2f1}\\
	&{}= \frac{1}{\Gamma(g-e)} \int_0^1 \mathrm{d}z\, z^{e-1} \bar{z}^{g-e-1} f_1 \left( \left.\begin{matrix}a;\,c,\,d\, \\ b\,\end{matrix} \right\rvert x, yz \right),
\label{eq:110f121viaf1}\\
	&{}= \Gamma\begin{bmatrix} c,\, b-c \\ a,\, b-a \end{bmatrix} \bar{x}^{b-a-c} \int_0^1 \mathrm{d}z\, \frac{ z^{b-c-1} \bar{z}^{c-1} }{ (1- x z)^{b-a} } {}_3f_2 \left( \left.\begin{matrix}a,\,d,\,e\, \\ b-c,\,g\, \end{matrix} \right\rvert yz \right),
\label{eq:110f121via3f2}
\end{align}
where $\Re b > \Re a > 0$ and, for \eqref{eq:110f121via3f2}, $\Re b > \Re c > 0$; $f_1(x,y) = \mathsf{f}^{1:1;1}_{1:0;0}(x,y)$ in eq.~\eqref{eq:AppellF1}. The equivalence of the first two representations can be easily proved by using the Euler integrals for $_2f_1$ \eqref{fwc:07.23.07.0001.01} and $f_1$ \eqref{fwc:07.36.07.0001.01}. The third representation can be obtained from the second one with the help of the following representation for $f_1$ \cite{brychkov2012some}:
\begin{align}
	f_1 \left( \left.\begin{matrix}a;\,c,\,d\, \\ b\,\end{matrix} \right\rvert x, y \right)
	=
	\bar{x}^{b-a-c} \int_0^1 \mathrm{d}z\, \frac{ z^{b-c-1} \bar{z}^{c-1} }{ (1- x z)^{b-a} }\, {}_2f_1 \left( \left.\begin{matrix}a,\,d\, \\ b-c\, \end{matrix} \right\rvert yz \right).
\end{align}
Eqs.~\eqref{eq:110f121via2f1}--\eqref{eq:110f121via3f2} allows us to obtain \eqref{eq:1111n-xb} readily by integrating \eqref{eq:1111n-fact} with $y^b$ over $y$ (one should simply change the integration variable to $z$).

Making the substitution $z \to \bar{z}$ in eq.~\eqref{eq:110f121via2f1}, setting $y=1$, and using the Kummer transformation \eqref{fwc:07.23.17.0061.01}, we obtain the following autotransformation property for $\mathsf{f}^{1:1;2}_{1:0;1}(x,1)$:
\begin{align}
	\bar{x}^c \mathsf{f}^{1:1;2}_{1:0;1} \left( \left.\begin{matrix}a&:&c&;&d,\,e\,\\ b&:& \text{---} &;&g\,\end{matrix} \right\rvert x, 1 \right)
	={}& \Gamma\begin{bmatrix} a,\, d+e-g,\, 1+g-d-e \\ b-a,\, g-d,\,g-e \end{bmatrix} 
\notag \\
	&{}\times \Biggl[ \mathsf{f}^{1:1;2}_{1:0;1} \left( \left.\begin{matrix} b-a+g-d-e&:&c&;&g-d,\,g-e\,\\ b+g-d-e&:& \text{---} &;&g-d-e+1\,\end{matrix} \right\rvert -\frac{x}{\bar{x}}, 1 \right)
\notag \\ \label{eq:110f121-au}
	& \hphantom{{}\times \Biggl[} {}- \mathsf{f}^{1:1;2}_{1:0;1} \left( \left.\begin{matrix}b-a&:&c&;&d,\,e\,\\ b&:& \text{---} &;&d+e-g+1\,\end{matrix} \right\rvert -\frac{x}{\bar{x}}, 1 \right) \Biggr].
\end{align}

Also, we can make the substitution $z \to \bar{z}$ in eq.~\eqref{eq:110f121via2f1}, expand the denominator in the integrand, and evaluate the integral of the series term by term with the help of the integral representation of $_3F_2$ \eqref{fwc:07.31.07.0001.01}. This gives
\begin{align}\label{eq:110f121-au2}
	\bar{x}^c \mathsf{f}^{1:1;2}_{1:0;1} \left( \left.\begin{matrix}a&:&c&;&d,\,e\,\\ b&:& \text{---} &;&g\,\end{matrix} \right\rvert x, y \right)
	= \frac{1}{\Gamma(b-a)} \mathsf{f}^{0:2;3}_{1:0;1} \left( \left.\begin{matrix}\text{---}&:&b-a,\,c&;&a,\,d,\,e\,\\ b&:& \text{---} &;&g\,\end{matrix} \right\rvert -\frac{x}{\bar{x}}, y \right).
\end{align}

In a number of cases $\mathsf{f}^{1:1;2}_{1:0;1}(x,1)$ reduces to simpler hypergeometric functions. One of them follows from eq.~\eqref{eq:110f121via2f1}---the integral over $z$ can be easily done in terms of $_3f_2$ if $a$ is equal to $g$ \cite{prudnikov2003-en-2.21.1.22}. Indeed, expanding the denominator of the integrand as a series in $-x/\bar{x}$, evaluating the integral over $z$, and using the well-known summation formula for $_2f_1(1)$ \eqref{fwc:07.23.03.0002.01}, we get
\begin{align}\label{eq:BMPred}
	\bar{x}^c \mathsf{f}^{1:1;2}_{1:0;1} \left( \left.\begin{matrix}a&:&c&;&d,\,e\,\\ b&:& \text{---} &;&a\,\end{matrix} \right\rvert x, 1 \right)
	= \Gamma\begin{bmatrix}d,\, e \\ b-a \end{bmatrix} {}_3f_2 \left( \left.\begin{matrix} b-d-e,\, c,\, b-a\, \\ b-d,\,b-e\,\end{matrix} \right\rvert -\frac{x}{\bar{x}} \right).
\end{align}

Another possibility of reduction important for us has been pointed out in our sketching the proof of eq.~\eqref{eq:KdFto3f2red1}. It is related to the reduction \eqref{fwc:07.36.03.0006.01} of $f_1$ to $_2f_1$ in the integral representation \eqref{eq:110f121viaf1} if $b=c+d$. We will not develop the consideration of section~\ref{subsec:3F2} here. It should be noted, however, that the reduction formula~\eqref{eq:f112101-red-1} can be derived (rather cumbersomely) in the same way---use the representation \eqref{eq:110f121viaf1}, write the Appell function as $_2f_1$ (see \eqref{fwc:07.36.03.0006.01}), transform $_2f_1$ with the help of \eqref{fwc:07.23.17.0061.01}, evaluate the resulting simple integrals, and apply the identity \eqref{fwc:07.27.17.0026.01} to $_3f_2(1/\bar{x})$.

Now, let us prove the reduction~\eqref{eq:f112101-red-2}. To this end, we represent the KdF function as a series with $_3f_2(1)$ in coefficients:
\begin{multline}
\mathsf f^{1:1;2}_{1:0;1} \left(  \left.\begin{matrix}1&: & b+\lambda-\dot{n}&;& n,\,\dot{n}+1 \\
																			 1+\lambda&:&\text{---}&;&2(\dot{n}+1)
														\end{matrix} \right\rvert \bar x, 1 \right)
 \\
=
\sum_{r \geqslant 0} \frac{\Gamma(b+\lambda-\dot{n}+r)}{r!} \bar x^r {}_3f_2 \left(  \left.\begin{matrix}1+r,\; n,\; \dot{n}+1 \\
																			 1+\lambda+r,\; 2(\dot{n}+1)
														\end{matrix} \right\rvert 1 \right).
\end{multline}
Then, we transform $_3f_2(1)$ with the help of the identity \eqref{fwc:07.27.17.0039.01}
\begin{align}
{}_3f_2 \left(  \left.\begin{matrix}1+r,\; \dot{n}+1,\; n\, \\
										1 + \lambda + r,\; 2 ( \dot{n}+1)\,
						\end{matrix} \right\rvert 1 \right)
=
{}&\Gamma\begin{bmatrix}1+r,\; -\dot{n},\; \lambda -\dot{n}-1,\; r-2\dot{n},\; \dot{n}+1,\; n \\
								\lambda,\; -2 \dot{n}-1,\; r-\dot{n}+1,\; r+\lambda-\dot{n},\; 2 ( \dot{n}+1)
\end{bmatrix}
\notag\\&{}+{}_3f_2 \left(  \left.\begin{matrix}r-2 \dot{n},\; \lambda-\dot{n}-1,\; -\dot{n}\, \\
													r+\lambda-2\dot{n},\; -2 \dot{n}\,
								\end{matrix} \right\rvert 1 \right).
\end{align}
This results in the following expression:
\begin{align}
\mathsf f^{1:1;2}_{1:0;1} & \left(  \left.\begin{matrix}1&: & b+\lambda-\dot{n}&;& n,\,\dot{n}+1 \\
																			 1+\lambda&:&\text{---}&;&2(\dot{n}+1)
														\end{matrix} \right\rvert \bar x, 1 \right)
\notag \\
={}&
\Gamma\begin{bmatrix}\dot{n}+1,\; -\dot{n},\; \lambda-\dot{n}-1,\; n \\
								-2 \dot{n}-1,\; 2 ( \dot{n}+1),\; \lambda
			\end{bmatrix}
{}_3f_2 \left(  \left.\begin{matrix}1,\; b+\lambda-\dot{n},\; -2\dot{n}\, \\
											1-\dot{n},\; \lambda-\dot{n}\,
								\end{matrix} \right\rvert \bar{x} \right)
\notag\\
&{}+\mathsf f^{1:1;2}_{1:0;1} \left(  \left.\begin{matrix}-2 \dot{n}&: & b+\lambda-\dot{n}&;& \lambda-\dot{n}-1,\; -\dot{n}\, \\
																			 \lambda-2\dot{n}&:&\text{---}&;& -2 \dot{n}\,
														\end{matrix} \right\rvert \bar x, 1 \right).
\end{align}
The last term in the above equation can be simplified by virtue of the reduction \eqref{eq:BMPred}, which leads us immediately to eq.~\eqref{eq:f112101-red-2}.

Finally, to derive the reduction of the third term in eq.~\eqref{eq:1111n-xb} for $b=0$, we can make use of the autotransformation property \eqref{eq:110f121-au}, which gives two terms---the Appell function $f_1(-x/\bar{x},1)$ and a new $\mathsf{f}^{1:1;2}_{1:0;1}(-x/\bar{x},1)$. Both terms can be reduced to series in one variable due to the reduction formulas \eqref{eq:BMPred} and \eqref{fwc:07.36.03.0003.01}. In the case of $b=0$, the fourth term in eq.~\eqref{eq:1111n-xb} is equal to the third one up to replacing $x$ by $\bar{x}$.

%========================================================================================
%========================================================================================
%========================================================================================

\section{\label{App:C}Listing of parameter arrays of hypergeometric functions}

Here, we list the arrays of the parameters of the two-row gamma and generalized Lauricella functions in eqs.~\eqref{eq:MM-xb} and \eqref{eq:MM_ab}:
\begin{gather}\label{eq:PQ}
\begin{aligned}
[P_1] &{}= \begin{bmatrix}
		n_{2, 4,\tilde 5},\, b-n_{2,\tilde 4,\tilde 5},\, \lambda,\, n_{3, 4,\tilde 5} \\
		1
	\end{bmatrix},
\qquad
[P_2] = \begin{bmatrix}
		n_{2, 4,\tilde 5},\, b-n_{2,\tilde 4,\tilde 5},\, \lambda,\, n_{1, 2,\tilde 5} \\
		1
	\end{bmatrix},
\\
[Q_1] &{}= \begin{bmatrix}\text{---}&:& n_{3,4,\tilde 5},\,n_{1,3,\tilde 5},\, n_5&;&\lambda,\,n_3,\,b-n_{\tilde 4}&;& b-n_{2,\tilde 4,\tilde 5},\,-n_{\tilde 5}\,\\
										(b-n_{\tilde 4, \tilde 5}:\beta_1),(n_{3,5}:\beta_2)&:&n_{1,2,3,4,\tilde 5,\tilde 5}&;& \text{---}&;&\text{---}\,\end{bmatrix},
\\
[Q_2] &{}= \begin{bmatrix}\text{---}&:& n_{1,2,\tilde 5},\,n_{1,3,\tilde 5},\, n_5&;&\lambda,\,n_1,\,-n_{\tilde 2}&;& b-n_{2,\tilde 4,\tilde 5},\,-n_{\tilde 5}\,\\
										(-n_{\tilde 2, \tilde 5}:\beta_1),(n_{1,5}:\beta_2)&:&n_{1,2,3,4,\tilde 5,\tilde 5}&;& \text{---}&;&\text{---}\,\end{bmatrix},
\end{aligned}
\\ \notag
[\beta_1] = [0,1,1], \qquad [\beta_2] = [1,1,0].
\end{gather}
The arrays $P'_i$ and $Q'_i$, $i=1,2$ in eq.~\eqref{eq:MM_ab} are obtained from $P_i$ and $Q_i$ by attaching two more parameters in the last columns:
\begin{align}\label{eq:P'Q'}
\begin{alignedat}{2}
[P_1']&{}= \left[ P_1\; \begin{matrix*}[l],-n_{1,\tilde 2,\tilde 5}\\,1\end{matrix*} \right],
&\qquad
[Q_1']&{}= \left[ Q_1\; \begin{matrix*}[l],a+b-n_{3,\tilde 4,\tilde 5}\\,a+b-n_{1,2,\tilde 3,\tilde 4,\tilde 5,\tilde 5}\end{matrix*} \right],
\\
[P_2']&{}= \left[ P_2\; \begin{matrix*}[l],a-n_{2, \tilde 3,\tilde 5}\\,1\end{matrix*} \right],
&
[Q_2']&{}= \left[ Q_2\; \begin{matrix*}[l],b-n_{1,\tilde 4,\tilde 5}\\,a+b-n_{1,2,\tilde 3,\tilde 4,\tilde 5,\tilde 5}\end{matrix*} \right].
\end{alignedat}
\end{align}

The parameters of eq.~\eqref{eq:g0y} read
\begin{gather}\label{app:ablist}
\begin{gathered}
\alpha_1 = \alpha_3 = \alpha_5 = \alpha_7 = \alpha_9 = -n_{\tilde 4},
\qquad
\alpha_2 = \alpha_4 = \alpha_6 = \alpha_8 = \alpha_{10} = -n_{3, \tilde 4, \tilde 5},
\\
\beta_1 = \beta_2 = -n_{\tilde 2},
\quad
\beta_3 = \beta_4 = -n_{1, \tilde 2, \tilde 5},
\quad
\beta_5 = \beta_7 = \beta_9 = n_4,
\quad
\beta_6 = \beta_8 = \beta_{10} = n_{3, 4, \tilde 5};
\end{gathered}
\end{gather}

\begin{gather}
[\Xi_0] = \begin{bmatrix}
		n_{3, \tilde 5},\, 1-n_{3, \tilde 5} \\
		n_1,\, n_3,\, n_5,\, -n_{1, \tilde 3, \tilde 5}
	\end{bmatrix},
\quad
[\Xi_6] = \begin{bmatrix}
		-n_{2, 3, \tilde 4, \tilde 5},\, 1+n_{2, 3, \tilde 4, \tilde 5},\, -n_{1, 2, 3, \tilde 4, \tilde 5, \tilde 5},\, 1+n_{1, 2, 3, \tilde 4, \tilde 5, \tilde 5} \\
		-n_{3, \tilde 4, \tilde 5},\, n_{3, 4, \tilde 5},\, 1-n_{3, 4, \tilde 5},\, 1+n_{3, \tilde 4, \tilde 5}
	\end{bmatrix},
\notag\\
\begin{alignedat}{2}
[\Xi_1] &{}= \begin{bmatrix}
		-n_{1, \tilde 5},\, n_{2, \tilde 4},\, 1+n_{1, \tilde 5},\, 1-n_{2, \tilde 4} \\
		n_2,\, n_4,\, 1-n_2,\, 1-n_4
	\end{bmatrix},
&
[\Xi_7] &{}= \begin{bmatrix}
		1 \\
		n_4,\, 1-n_4
	\end{bmatrix},
\\
[\Xi_2] &{}= \begin{bmatrix}
		-n_{1, \tilde 5},\, 1+n_{1, \tilde 5},\, n_{2, 3, \tilde 4, \tilde 5},\, 1-n_{2, 3, \tilde 4, \tilde 5} \\
		n_2,\, 1-n_2,\, n_{3, 4, \tilde 5},\, 1-n_{3, 4, \tilde 5}
	\end{bmatrix},
&
[\Xi_8] &{}= \begin{bmatrix}
		1 \\
		n_{3, 4, \tilde 5},\, 1-n_{3, 4, \tilde 5}
	\end{bmatrix},
\\
[\Xi_3] &{}= \begin{bmatrix}
		n_{1, \tilde 5},\, 1-n_{1, \tilde 5},\, n_{1, 2, \tilde 4, \tilde 5},\, 1-n_{1, 2, \tilde 4, \tilde 5} \\
		n_4,\, 1-n_4,\, n_{1, 2, \tilde 5},\, 1-n_{1, 2, \tilde 5}
	\end{bmatrix},
&
[\Xi_9] &{}= \begin{bmatrix}
		-n_{2, \tilde 4},\, -n_{1, 2, \tilde 4, \tilde 5} \\
		n_4,\, 1-n_4,\, -n_{\tilde 4}
	\end{bmatrix},
\\
[\Xi_4] &{}= \begin{bmatrix}
		n_{1, \tilde 5},\, 1-n_{1, \tilde 5},\, n_{1, 2, 3, \tilde 4, \tilde 5, \tilde 5},\, 1-n_{1, 2, 3, \tilde 4, \tilde 5, \tilde 5} \\
		n_{1, 2, \tilde 5},\, n_{3, 4, \tilde 5},\, 1-n_{1, 2, \tilde 5},\, 1-n_{3, 4, \tilde 5}
	\end{bmatrix},
& \qquad\;
[\Xi_{10}] &{}= \begin{bmatrix}
		-n_{2, 3, \tilde 4, \tilde 5},\, -n_{1, 2, 3, \tilde 4, \tilde 5, \tilde 5} \\
		-n_{3, \tilde 4, \tilde 5},\, n_{3, 4, \tilde 5},\, 1-n_{3, 4, \tilde 5}
	\end{bmatrix};
\\
[\Xi_5] &{}= \begin{bmatrix}
		-n_{2, \tilde 4},\, 1+n_{2, \tilde 4},\, -n_{1, 2, \tilde 4, \tilde 5},\, 1+n_{1, 2, \tilde 4, \tilde 5} \\
		n_4,\, 1-n_4,\, -n_{\tilde 4},\, 1+n_{\tilde 4}
	\end{bmatrix},
&&
\end{alignedat}
\label{eq:Xilist}
\end{gather}

\begin{gather}
\begin{aligned}
[\Phi_1] &{}= \begin{bmatrix}
					n_5,\, n_{1, 3, \tilde 5} &:& 1,\, 1-n_2 &;& 1-n_4 \\
					1-n_{2, \tilde 4} &:& 1,\, 1+n_{1, \tilde 5} &;& 1+n_{3, \tilde 5}
				\end{bmatrix},
\;\;\;
[\Phi_2] = \begin{bmatrix}n_1,\, -n_{\tilde 3} &:& 1,\, 1-n_2 &;& 1-n_{3, 4, \tilde 5} \\
									1-n_{2, 3, \tilde 4, \tilde 5} &:& 1,\, 1+n_{1,\, \tilde 5} &;& 1-n_{3,\, \tilde 5}
				\end{bmatrix},
\\
[\Phi_3] &{}= \begin{bmatrix}n_3,\, -n_{\tilde 1}  &:& 1,\, 1-n_{1, 2, \tilde 5} &;& 1-n_4 \\
									1-n_{1, 2, \tilde 4, \tilde 5} &:& 1,\, 1-n_{1, \tilde 5} &;& 1+n_{3, \tilde 5}
				\end{bmatrix},
\\
[\Phi_4] &{}=  \begin{bmatrix}-n_{\tilde 5},\,-n_{1, \tilde 3, \tilde 5} &:& 1,\, 1-n_{1, 2, \tilde 5} &;& 1-n_{3, 4, \tilde 5} \\
									1-n_{1, 2, 3, \tilde 4, \tilde 5, \tilde 5}&:& 1,\, 1-n_{1, \tilde 5} &;& 1-n_{3, \tilde 5}
				\end{bmatrix},
\\
[\Phi_5] &{}= \begin{bmatrix}n_{2, 4, \tilde 5},\, n_{1, 2, 3, \tilde 4, \tilde 5} &:& 1,\, 1+n_{\tilde 4} &;& 1-n_4 \\
									1 &:& 1+n_{2, \tilde 4},\, 1+n_{1, 2, \tilde 4, \tilde 5} &;& 1+n_{3, \tilde 5}
				\end{bmatrix},
\\
[\Phi_6] &{}= \begin{bmatrix}n_{2, 4, \tilde 5},\, n_{1, 2, 3, \tilde 4, \tilde 5} &:& 1,\, 1+n_{3, \tilde 4, \tilde 5} &;& 1-n_{3, 4, \tilde 5} \\
									1 &:& 1+n_{2, 3, \tilde 4, \tilde 5},\, 1+n_{1, 2, 3, \tilde 4, \tilde 5, \tilde 5} &;& 1-n_{3, \tilde 5}
				\end{bmatrix},
\\
[\Phi_7] &{}= \begin{bmatrix}1-n_4 &:& n_{2, 4, \tilde 5},\, n_{1, 2, 3, \tilde 4, \tilde 5}  &;& 1,\, -n_{2, \tilde 4},\, -n_{1, 2, \tilde 4, \tilde 5} \\
									1,\, 1+n_{3, \tilde 5} &:& \text{---} &;& -n_{\tilde 4}
				\end{bmatrix},
\notag\\
[\Phi_8] &{}= \begin{bmatrix}1-n_{3, 4, \tilde 5} &:& n_{2, 4, \tilde 5},\, n_{1, 2, 3, \tilde 4, \tilde 5} &;& 1,\, -n_{2, 3, \tilde 4, \tilde 5},\, -n_{1, 2, 3, \tilde 4, \tilde 5, \tilde 5} \\
									1,\, 1-n_{3, \tilde 5} &:& \text{---} &;& -n_{3, \tilde 4, \tilde 5}
				\end{bmatrix},
\notag\\
[\Phi_9] &{}= \begin{bmatrix}1-n_4,\, n_{2, 4, \tilde 5},\, n_{1, 2, 3, \tilde 4, \tilde 5} \\
									1,\, 1+n_{3, \tilde 5}
				\end{bmatrix},
\qquad
[\Phi_{10}] = \begin{bmatrix}n_{2, 4, \tilde 5},\, 1-n_{3, 4, \tilde 5},\, n_{1, 2, 3, \tilde 4, \tilde 5} \\
									1,\, 1-n_{3, \tilde 5}
				\end{bmatrix}.
\label{app:Philist}
\end{aligned}
\end{gather}

The arrays of the parameters $\Xi_k'$ and $\Phi_k'$ in eq.~\eqref{eq:g0b} are as follows:
\begin{gather}
\Xi_k' = \Xi_k, \quad k=0,\dots, 6,
\notag\\
\begin{aligned}
[\Xi_7'] &{}= [\Xi_8'] = \begin{bmatrix}
		1 \\
		1
	\end{bmatrix},
\qquad
[\Xi_9'] = \begin{bmatrix}
		-n_{2, \tilde 4},\, -n_{1, 2, \tilde 4, \tilde 5} \\
		1-n_4,\, -n_{\tilde 4}
	\end{bmatrix},
\qquad
[\Xi_{10}'] = \begin{bmatrix}
		-n_{2, 3, \tilde 4, \tilde 5},\, -n_{1, 2, 3, \tilde 4, \tilde 5, \tilde 5} \\
		-n_{3, \tilde 4, \tilde 5},\, 1-n_{3, 4, \tilde 5}
	\end{bmatrix},
\end{aligned}\label{app:Xi'list}
\end{gather}
where $\Xi_k$, $k=0,\dots, 10$ are given in eq.~\eqref{eq:Xilist};

\begin{gather}
\begin{aligned}
[\Phi_1'] &{}= \begin{bmatrix}
					n_5,\, n_{1, 3, \tilde 5} &:& 1,\, 1-n_2,\, -n_{\tilde 2} &;& 1-n_4,\, b-n_{\tilde 4} \\
					1-n_{2, \tilde 4},\, b-n_{\tilde 2, \tilde 4} &:& 1,\, 1+n_{1, \tilde 5} &;& 1+n_{3, \tilde 5}
				\end{bmatrix},
\\
[\Phi_2'] &{}= \begin{bmatrix}n_1,\, -n_{\tilde 3} &:& 1,\, 1-n_2,\, -n_{\tilde 2} &;& 1-n_{3, 4, \tilde 5},\, b-n_{3, \tilde 4, \tilde 5} \\
									1-n_{2, 3, \tilde 4, \tilde 5},\, b-n_{2, \tilde 3, \tilde 4, \tilde 5} &:& 1,\, 1+n_{1,\, \tilde 5} &;& 1-n_{3,\, \tilde 5}
				\end{bmatrix},
\\
[\Phi_3'] &{}= \begin{bmatrix}n_3,\, -n_{\tilde 1}  &:& 1-n_{1, 2, \tilde 5},\,-n_{1, \tilde 2, \tilde 5} &;& 1-n_4,\, b-n_{\tilde 4} \\
									1-n_{1, 2, \tilde 4, \tilde 5},\, b-n_{1, \tilde 2, \tilde 4, \tilde 5} &:&1-n_{1, \tilde 5} &;& 1+n_{3, \tilde 5}
				\end{bmatrix},
\\
[\Phi_4'] &{}=  \begin{bmatrix}-n_{\tilde 5},\,-n_{1, \tilde 3, \tilde 5} &:& 1-n_{1, 2, \tilde 5},\, -n_{1, \tilde 2, \tilde 5} &;& 1-n_{3, 4, \tilde 5},\, b-n_{3, \tilde 4, \tilde 5} \\
									1-n_{1, 2, 3, \tilde 4, \tilde 5, \tilde 5},\, b-n_{1, \tilde 2, \tilde 3, \tilde 4, \tilde 5}&:& 1-n_{1, \tilde 5} &;& 1-n_{3, \tilde 5}
				\end{bmatrix},
\\
[\Phi_5'] &{}= \begin{bmatrix}n_{2, 4, \tilde 5},\, n_{1, 2, 3, \tilde 4, \tilde 5} &:& 1,\, 1+n_{\tilde 4},\, n_4 &;& 1-n_4,\, b-n_{\tilde 4} \\
									1, b+D/2 &:& 1+n_{2, \tilde 4},\, 1+n_{1, 2, \tilde 4, \tilde 5} &;& 1+n_{3, \tilde 5}
				\end{bmatrix},
\\
[\Phi_6'] &{}= \begin{bmatrix}n_{2, 4, \tilde 5},\, n_{1, 2, 3, \tilde 4, \tilde 5} &:& 1,\, 1+n_{3, \tilde 4, \tilde 5},\, n_{3, 4, \tilde 5} &;& 1-n_{3, 4, \tilde 5},\, b-n_{3, \tilde 4, \tilde 5} \\
									1,\, b+D/2 &:& 1+n_{2, 3, \tilde 4, \tilde 5},\, 1+n_{1, 2, 3, \tilde 4, \tilde 5, \tilde 5} &;& 1-n_{3, \tilde 5}
				\end{bmatrix},
\\
[\Phi_7'] &{}= \begin{bmatrix}1-n_4,\, -n_{\tilde 4} &:& 1,\, -n_{2, \tilde 4},\, -n_{1, 2, \tilde 4, \tilde 5} &;& n_{2, 4, \tilde 5},\, n_{1, 2, 3, \tilde 4, \tilde 5} \\
									1,\, 1+n_{3, \tilde 5} &:& -n_{\tilde 4},\, 1-n_4 &;& b+D/2
				\end{bmatrix},
\\
[\Phi_8'] &{}= \begin{bmatrix}1-n_{3, 4, \tilde 5},\, b-n_{3, \tilde 4, \tilde 5} &:& 1,\, -n_{2, 3, \tilde 4, \tilde 5},\, -n_{1, 2, 3, \tilde 4, \tilde 5, \tilde 5} &;& n_{2, 4, \tilde 5},\, n_{1, 2, 3, \tilde 4, \tilde 5} \\
									1,\, 1-n_{3, \tilde 5} &:& -n_{3, \tilde 4, \tilde 5},\, 1-n_{3, 4, \tilde 5} &;& b+D/2
				\end{bmatrix}.
\\
[\Phi_9'] &{}= \begin{bmatrix}1-n_4,\, n_{2, 4, \tilde 5},\, n_{1, 2, 3, \tilde 4, \tilde 5},\, b-n_{\tilde 4} \\
									1,\, 1+n_{3, \tilde 5},\, b+D/2
				\end{bmatrix},
\,\,\,
[\Phi_{10}'] = \begin{bmatrix}n_{2, 4, \tilde 5},\, 1-n_{3, 4, \tilde 5},\, n_{1, 2, 3, \tilde 4, \tilde 5},\, b-n_{3, \tilde 4, \tilde 5} \\
									1,\, 1-n_{3, \tilde 5},\, b+D/2
				\end{bmatrix}.
\end{aligned}
\label{app:Phi'list}
\end{gather}

%===================================================
%===================================================
%===================================================

\providecommand{\href}[2]{#2}\begingroup\raggedright\endgroup

\end{document}